\documentclass[%
 reprint,
superscriptaddress,
 amsmath,amssymb,
 aps,
floatfix,
]{revtex4-2}

\usepackage{graphicx}
\usepackage{dcolumn}
\usepackage{bm}
\usepackage{hyperref}
\usepackage{float}
\usepackage{comment}
\usepackage[version=3]{mhchem}

\usepackage[normalem]{ulem}

\def\*#1{\mathbf{#1}}

\begin{document}


\title{Electronic properties of MoSe$_2$ nanowrinkles}

\author{Stefan Velja}
 \affiliation{Institute of Physics, Carl von Ossietzky Universit\"at Oldenburg, 26129 Oldenburg, Germany}

\author{Jannis Krumland}
 \affiliation{Department of Physics and IRIS Adlershof, Humboldt-Universit\"at zu Berlin, 12489 Berlin, Germany}
\affiliation{Institute of Physics, Carl von Ossietzky Universit\"at Oldenburg, 26129 Oldenburg, Germany}

\author{Caterina Cocchi}
 \affiliation{Institute of Physics, Carl von Ossietzky Universit\"at Oldenburg, 26129 Oldenburg, Germany}
 \affiliation{Center for Nanoscale Dynamics (CeNaD), Carl von Ossietzky Universit\"at Oldenburg, 26129 Oldenburg, Germany}
 \affiliation{Department of Physics and IRIS Adlershof, Humboldt-Universit\"at zu Berlin, 12489 Berlin, Germany}

\date{\today}

\begin{abstract}
Mechanical deformations, either spontaneously occurring during sample preparation or purposely induced in their nanoscale manipulation, drastically affect the electronic and optical properties of transition metal dichalcogenide monolayers. In this first-principles work based on density-functional theory, we shed light on the interplay among strain, curvature, and electronic structure of \ce{MoSe2} nanowrinkles. We analyze their structural properties highlighting the effects of coexisting local domains of tensile and compressive strain in the same system. By contrasting the band structures of the nanowrinkles against counterparts obtained for flat monolayers subject to the same amount of strain, we clarify that the specific features of the former, such as the moderate variation of the band-gap size and its persisting direct nature, are ruled by curvature rather than strain. The analysis of the wave-function distribution indicates strain-dependent localization of the frontier states in the conduction region while in the valence the sensitivity to strain is much less pronounced. The discussion about transport properties, based on the inspection of the effective masses, reveals excellent perspectives for these systems as active components for (opto)electronic devices. 
\end{abstract}

\maketitle


\section{\label{Intro}Introduction}
The exceptional electronic and optical properties of transition metal dichalcogenide (TMDC) monolayers have made them promising candidates for integrated electronics, optics, and photonics~\citep{shi+15csr,brar+18csr,ko+20jpd,mait+20natp,ghol+23jpp}.
The mechanical flexibility of these two-dimensional sheets additionally enables the manipulation of their intrinsic features through deposition on nanostructured substrates~\cite{jeon+17am,part+21natcom,li+22nano,kaya+23nl} where significant deformations may occur giving rise to peculiar topologies~\cite{blundo2020evidence,zhan+21im,balg+22ieeejqe}, such as nanobubbles~\citep{shep+17_2DM,blun+21prl,kim+22nano} or nanowrinkles~\cite{dirn+21sa}.
These nanostructures offer the opportunity to further tune the characteristics of the TMDCs, especially concerning exciton manipulation~\cite{darl+20natn,long+20ns,jian-patc22ns,lee+23pccp,ghos+23am} and quantum emission~\cite{part+21natcom,kim+22nano}.
However, the complex interplay between structural characteristics, including strain and curvature, and electronic and optical response~\cite{yu+16nl,carm+19nl,jian-patc22ns, wang2020strain,yu2023strain} does not facilitate the identification of a straightforward rationale to engineer their properties.

Conventional TMDCs formed by Mo or W and S or Se share most of their electronic and optical properties. In the monolayer form, they are all direct-band-gap semiconductors with exceptional photoluminescence~\cite{teby+20nano}. In response to shear and strain, their electronic structure exhibits a qualitatively similar response~\cite{mani+16ssc}
Among these materials, \ce{MoSe2} has been regarded with particular attention in several application areas, ranging from sensing~\cite{zhao+22ccr,kumar2024nanoparticles} to energy harvesting~\cite{wang+15jpcc,bissett2016comparison,shen+19ns}, from photodetection~\cite{chan+14nano,blau+19acsphotonics} to spin filtering~\cite{li+18cpl,zhan+19nano,khan+20aplmater}. This large spectrum of technological perspectives is aided by the peculiar tunability of \ce{MoSe2} with respect to the substrate~\cite{huang2024substrate}.
Remarkably, interest in this material is present also in industrial research~\cite{patent_detector,patent_transistor,patent_battery,patent_conversion,patent_energy}. Thanks to its lower predicted Young's modulus in comparison with the other conventional TMDCs~\cite{deng+18pe}, it is expected that \ce{MoSe2} is expected to be the most flexible member of this family of materials. From the computational perspective, it is more appealing than its W-based counterpart due to the smaller size of the metallic species.

First-principles methods such as density functional theory (DFT) are ideally suited to describe and predict the electronic features of complex materials. 
In particular, they enable systematic studies of structure-property relationships without the input of experimental data. 
This is very important in the case of rippled TMDCs, where the large number of degrees of freedom in play makes the effects of curvature, strain, and local deformations particularly hard to disentangle~\cite{yu+16nl,jian-patc22ns}.
Moreover, in contrast to classical models, which are useful to reveal the mechanical characteristics of peculiar low-dimensional topologies~\cite{blun+21prl,blun+22nl} and to connect them with macroscopic experimental variables, a quantum-mechanical description of the electronic structure is able to provide microscopic insight into these materials~\cite{yu+16nl,jian-patc22ns,carm+19nl,ocho+17prl}.

In this work, we investigate from first principles the electronic properties of \ce{MoSe2} nanowrinkles with varying curvature and strain. 
The modeled systems are of nanometer size and thus much smaller than experimental equivalents that have been fabricated on the micrometer scale on nanoengineered substrates and/or under specific mechanical constraints~\cite{castellanos2013local,liu+20nl,dirn+21sa,wang+21nl,liu+22acsami,ghos+23am}. Nonetheless, they offer a suitable platform to investigate the electronic behavior of such systems on a pure quantum mechanical level. We analyze the structural characteristics of the simulated nanowrinkles by discussing the coexistence of local domains of tensile and compressive strain in the same sample. We explain their role in the electronic properties examined through band structure plots and the analysis of wave-function distributions at the frontier states. By comparing the results obtained for the nanowrinkles with those of flat monolayers subject to the same amount of strain, we reveal the role of curvature in preserving the size and the direct nature of the band gap regardless of the amount of deformation. We calculate effective masses associated with the frontier states and in the conduction region, we find different trends depending on the strain value. These results not only offer important insight into the fundamental properties of TMDC nanowrinkles but also demonstrate the exceptional perspectives of these materials, when subject to controlled mechanical modulation, as efficient active components for optoelectronic and photonic devices.

\section{\label{Methods}Methodology}

\subsection{Theoretical background}
\label{ss:theory}

The results presented in this work are obtained in the framework of DFT~\cite{hohenberg1964inhomogeneous}, implemented in the Kohn-Sham (KS) scheme~\cite{kohn1965self}.
Electron energies $\varepsilon_i$ and wave-functions $\psi_i$ are calculated from the solution of the Schr\"odinger--like KS equation
\begin{equation}
    \left(-\frac{\hbar^2}{2m_\mathrm{e}}\nabla^2 + v_{\mathrm{eff}} -\varepsilon_i\right) \psi_i(\*r) = 0,
\end{equation}
where the effective potential $v_{\mathrm{eff}}$ is given by the sum of the external potential $v_{\mathrm{ext}}$, accounting for the electron-nuclear interaction, of the Hartree potential $v_{\mathrm{H}}$, and of the exchange-correlation potential $v_{\mathrm{xc}}$.
The exact form of the last term is unknown and thus must be approximated.
Local and semi-local $v_{\mathrm{xc}}$ are known to severely underestimate the band-gap of semiconductors~\cite{burk12jcp} but still offer a reliable and computationally efficient tool to evaluate the electronic properties of these systems on a qualitative level. 
It is worth stressing that these calculations are performed at zero temperature and hence do not provide any information about the thermodynamic stability of the investigated systems. However, given the significantly larger size of TMDC nanowrinkles fabricated so far~\cite{castellanos2013local,dirn+21sa,wang+21nl,liu+22acsami,ghos+23am}, this aspect cannot be properly assessed with the considered modeled structures and thus goes beyond the scope of the present work. 

To map the band structures of the nanowrinkles from the orthorhombic representation adopted in the calculations onto the hexagonal Brillouin zone (BZ) of the pristine monolayer, we adopted an unfolding procedure~\cite{popescu_zunger,boyk-klim05prb,boyk+07jpcm} implemented in-house~\cite{krumland2021conditions,krum-cocc23pssa}. To this end, we introduce the spectral function 
\begin{equation}
W(\*k, E)=\sum_n\sum_{\substack{\sigma=\uparrow,\downarrow \\ \*g\in r}}|c_{n \sigma,\*{k'}}(\*G_0+\*g)|^2\delta(E-\varepsilon_n(\*{k'}))
\label{eq:spectral}
\end{equation}
to assign the corresponding weight to each state in the target representation.
In Eq.~\eqref{eq:spectral}, $\varepsilon_n(\*{k'})$ is the energy of band $n$ at the point $\*{k'}$ of the supercell. The wave vector in the BZ of the hexagonal cell, $\*k,$ is related to $\*{k'}$ via a reciprocal lattice (RL) vector $\*G_0$ of the supercell such that $\*k=\*{k'}+\*G_0$. The dummy variable $\*g$ runs over the reciprocal lattice vectors pertaining to the primitive cell, while $\sigma$ caters to the spin-up and spin-down channels. The coefficients $c_{n \sigma,\*{k'}}$ indexed by these variables are defined through the plane-wave representation of corresponding spinor states 
\begin{equation}
|n\*{k'}\rangle = \sum_{\sigma=\uparrow,\downarrow} \left( \sum_{\*G\in R}c_{n \sigma,\*{k'}}(\*G) |\*{k'}+\*G\rangle \right) \otimes |\sigma \rangle,
\label{eq:spinor}
\end{equation}
where $\*G$ sums over all the RL vectors of the supercell. Further details about this procedure are reported in the SI, see Fig.~S2.

To evaluate effective masses, we approximate the highest-occupied and the lowest-unoccupied bands as parabolic, such that the energy dispersion can be calculated adopting the usual free-particle approximation:
\begin{equation}
E \approx \frac{\hbar^2 k^2}{2m^*},
\end{equation}
where the effective mass $m^*$ replaces the mass of the free electron.
The maximum or minimum of the parabola is assumed to be at the $\Gamma$-point, where the effective masses are computed. The effective mass of a given electronic state is extracted from parabolic fits of the first 4 data points of the frontier states along the $\Gamma$-X path in the orthorhombic BZ.

\subsection{Computational details}

All DFT calculations presented in this work were performed using Quantum ESPRESSO~\cite{giannozzi2009quantum, giannozzi2017advanced}, with the Perdew-Burke-Ernzerhof (PBE) approximation~\cite{perdew1996generalized} for the $v_{\mathrm{xc}}$ and SG15 ONCV pseudopotentials~\cite{schlipf2015optimization} including spin-orbit coupling (SOC). The kinetic energy cutoffs for the wave functions and the charge density are set to 60~Ry and 240~Ry, respectively. During structural optimization, the interatomic forces are minimized with a threshold of $10^{-5}$~Ha/bohr.  
The electron convergence threshold is set to $10^{-9}$~Ry in all calculations.
Along the non-periodic direction $z$, a vacuum layer with a thickness of 20~\AA{} is included in the unit cell (UC) to avoid spurious couplings between the replicas upon the application of periodic boundary conditions in all directions.

\section{\label{Results}Results and discussion}

\subsection{Structural Properties}
\label{ss:model} 

\begin{figure}[h!]
    \centering
    \includegraphics[width=0.45\textwidth]{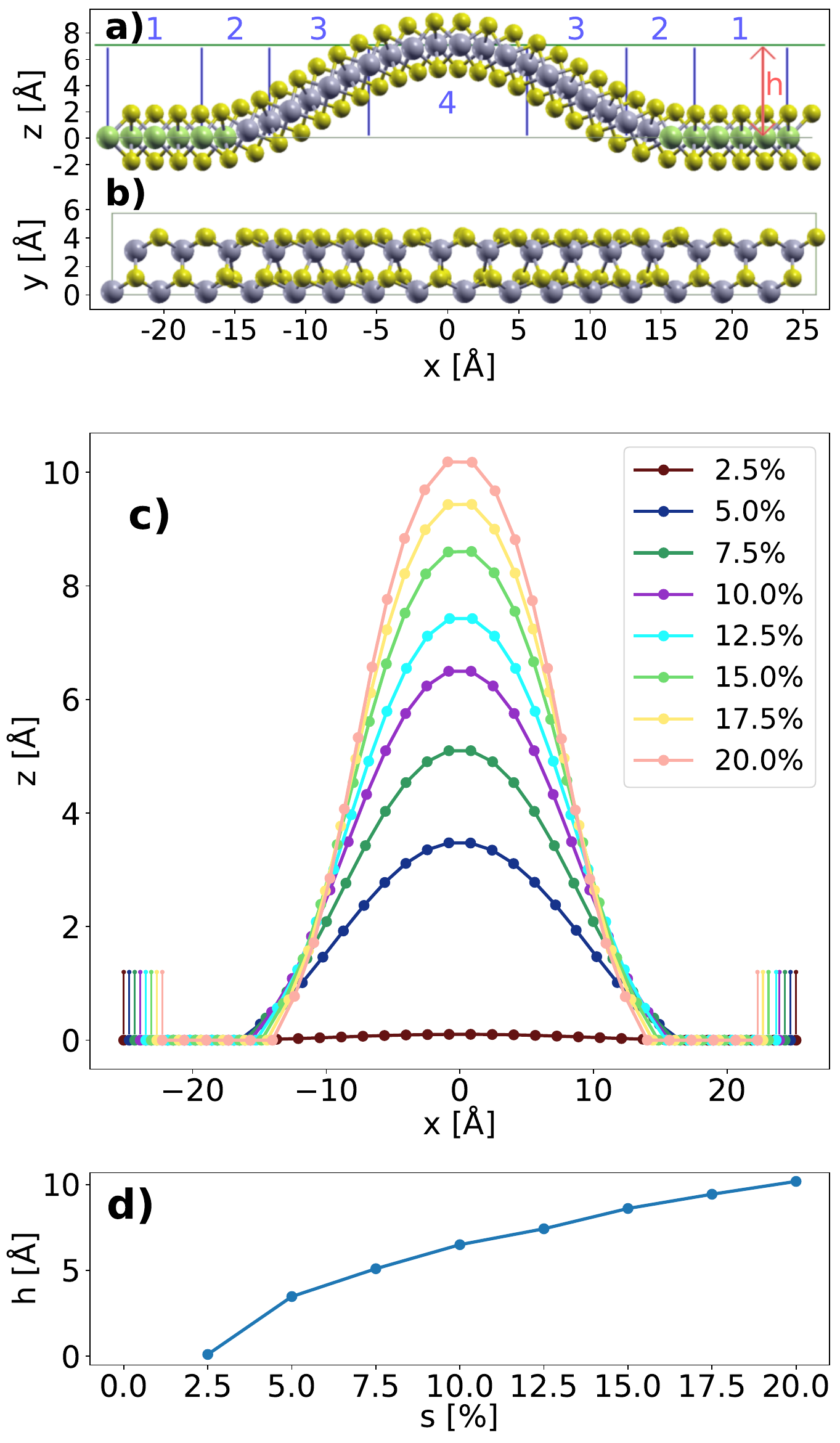}
    \caption{Ball-and-stick representation of a \ce{MoSe2} nanowrinkle with strain $s=10\%$, a) side view and b) top view. Mo atoms are visualized in grey and Se atoms in yellow. The atoms held fixed during structural optimization are highlighted in green in panel a). The height of the nanowrinkle $h$ is marked by a red arrow. We label the segments of the nanowrinkle (boundaries marked in blue) as follows: (1) periphery, (2) base, (3) wing, and (4) peak. c) $z$- vs. $x$-coordinates of the Mo atoms in all considered nanowrinkles represented by the colored dots; the legend refers to the corresponding strain $s$. Vertical bars emphasize the positions of the first and last Mo atoms of each nanowrinkle. d) Height $h$ of the considered nanowrinkles as a function of strain $s$.
    }
    \label{fig:example}
\end{figure}

To construct the \ce{MoSe2} nanowrinkles considered in this work, we start from the orthorhombic unit cell (OUC) of the material including two Mo and four Se atoms and with lattice parameters $a_0=3.30$~\AA{} along $x$ and $b = \sqrt{3} a_0 \approx 5.71$~\AA{} along $y$ (see Supplementary Material, Fig.~S1). Here, $a_0$ is the lattice constant of the \ce{MoSe2} monolayer in its conventional hexagonal UC obtained in this work through DFT optimization; the reported value is in agreement with the existing literature ~\cite{nepal2019first, ma2011electronic}. 
In a supercell including 16 replicas of the above-mentioned OUC, the central 10 represent the actual wrinkle, while the remaining 6 OUC in the lateral regions [further referred to as ``periphery"; see Fig.~\ref{fig:example}a)] are meant to isolate wrinkles from their replicas in neighboring unit cells. 
The periphery has a total length of $6 a_0 = 19.79$~\AA{} and the Mo atoms therein, marked in green in Fig.~\ref{fig:example}a), are kept fixed during optimization while all the other atoms are allowed to relax. We induce the characteristic curvature of the nanowrinkle by applying a certain amount of uniaxial strain $s$ along the $x$-axis to the central part of the supercell. Thus, the lattice constant $a$ of the supercell depends on $s$ as 
\begin{equation}
a = [6 + 10 \cdot (1-s)] a_0,
\label{eq:latconst}    
\end{equation}
where the factor of $6$ corresponds to the OUC at the periphery while $10 \cdot (1-s)$ caters for the 10 central OUC, strained by $s$. In practice, this is achieved by fixing the Mo atoms at the periphery in accordance with Eq.~\eqref{eq:latconst} and giving an initial guess for the position of the remaining atoms, which are then allowed to relax into the nanowrinkles we examine.

We create nanowrinkles with values of $s$ ranging from 2.5\% to 20\%, with increments of 2.5\%. According to the amount of applied strain, $a$ varies from 46.18~\AA{} ($s=20\%$) to 51.96~\AA{} ($s=2.5\%$).
The deformation induced by the strain can be quantified by introducing a ``height parameter'' $h$ defined as the difference between the $z$-coordinate of the uppermost Mo atom and the Mo atoms kept fixed in the lateral regions and assumed to be at 0~\AA{} [see Fig.~\ref{fig:example}a) and for a top view Fig.~\ref{fig:example}b)].
When the smallest strain $s=2.5\%$ is applied, the nanostructure remains almost completely flat and $h$ is equal to 0.1~\AA, see Fig.~\ref{fig:example}c).
Under higher strain, actual nanowrinkles are formed, and $h$ ranges from approximately 4~\AA{} for $s=5\%$ to 10~\AA{} when $s=20\%$.
By plotting $h$ as a function of $s$, the approximately linear behavior of the height \textit{vs.} strain of the nanowrinkle is apparent [see Fig.~\ref{fig:example}d)].
Expectedly, the result obtained for $s=2.5\%$ does not follow this trend, as the structure is substantially flat.

\begin{figure}
    \centering
    \includegraphics[width=0.5\textwidth-9pt]{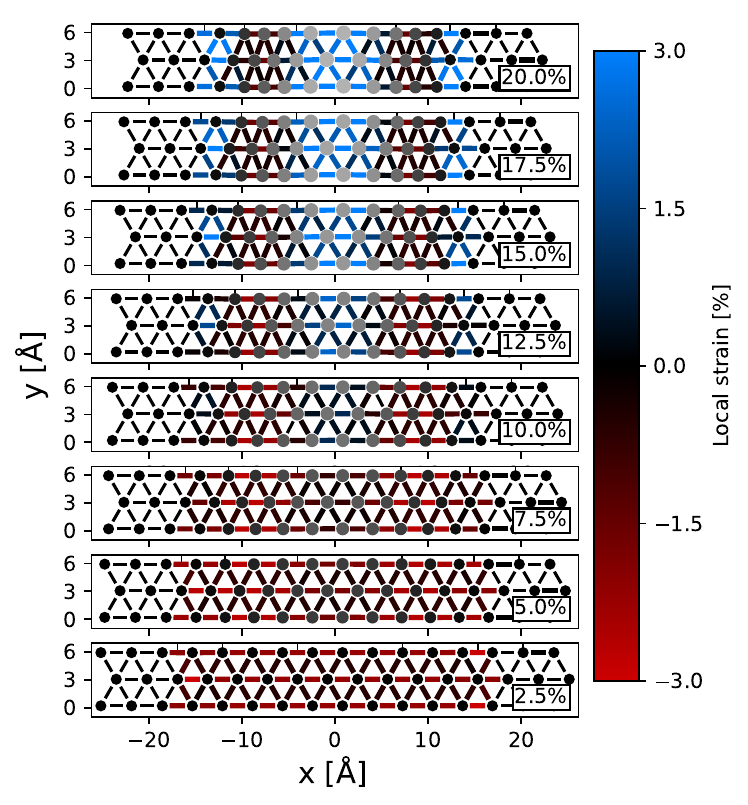}
    \caption{Map of local strain for all considered nanowrinkles at varying values of $s$ reported in the insets. The dots mark the positions of Mo atoms depicted in different shades of grey indicating the value of their $z$ coordinate: black means $z=0$~\AA{} while light gray $z=10$~\AA{}. The bars connecting the dots highlight the Mo-Mo distances and are colored according to the bar on the side: positive and negative percentage values indicate increments or reductions with respect to the reference value $a_0 = 3.30$~\AA{} in a flat and unstrained \ce{MoSe2} monolayer. The color bar is limited to 3\% from the upper end for clarity of $s\in\{7.5\%, 10\%, 12.5\%\}$, which are discussed in Subsection~\ref{ss:eff_mass}. Ticks on the upper x-axis denote the segments of the wrinkle, introduced in Fig.~\ref{fig:example}a).}
    \label{fig:local strain}
\end{figure}

To better understand the effects of strain on the structural properties of the nanowrinkles, we compare the distances between neighboring Mo atoms in the considered systems with respect to their distance in a pristine \ce{MoSe2} monolayer, which is equal to the lattice parameter $a_0 = 3.3$~\AA{} of its hexagonal UC.
We visualize the resulting values in Fig.~\ref{fig:local strain} by adopting a color code to differentiate among compression (negative strain values, in red), extension (positive strain values, in blue), and no strain (black). The dots in Fig.~\ref{fig:local strain} are color-coded according to the $z$-coordinate of the respective Mo atoms, ranging from $z=0$~\AA{} (black) to $z=10$~\AA{} (light gray).
Inspecting Fig.~\ref{fig:local strain} from bottom to top, we notice that upon the lowest strain, $s=2.5\%$, the distances between neighboring Mo atoms with the same $y$-coordinate are reduced by $2\%-3\%$ compared to their counterparts in the unstrained, flat monolayer. These values are not significantly different from $2.5\%$, which is expected given the flatness of this particular configuration.
At increasing values of strain, $s=5\%$ and $s=7.5\%$, we notice a more significant reduction in the Mo-Mo distances on the order of 3\% around the wings of the nanowrinkle. This aligns with the observation that significant system curvature appears for $s=5\%$, but not for $2.5\%$.
At the peak, on the other hand, Mo-Mo separations tend to expand slightly, although this behavior is hardly visible in the color scale adopted in Fig.~\ref{fig:local strain}.
In the nanowrikles with $s=10\%$ and $s=12.5\%$, the same trends persist: the Mo-Mo distances on the slope further decrease while those at the peak expand, becoming larger than $a_0$.
For $s=15\%$, we find not only a visible extension of the Mo-Mo separation at the peak but also an increment of the same quantity at both bases, exhibiting central reflection symmetry with respect to the middle of the wrinkle.
In the most strained nanowrinkles ($s=17.5\%$ and $s=20\%$), there is an alternation of extension and compression domains of Mo-Mo distances between the straight and curved parts of the system. This seems to be a primarily geometrical feature, as the Se atoms at the inner sides of the curves are more closely packed, giving rise to stronger mutual repulsion that influences nearby Mo atoms as well.
In the nanowrinkle with $s=20\%$, some Mo-Mo separations are almost 9\% larger than in the flat and pristine monolayer. 

\subsection{Band structures}
\label{ss:bs}

\begin{figure}[h!]
\centering
\includegraphics[width=0.48\textwidth]{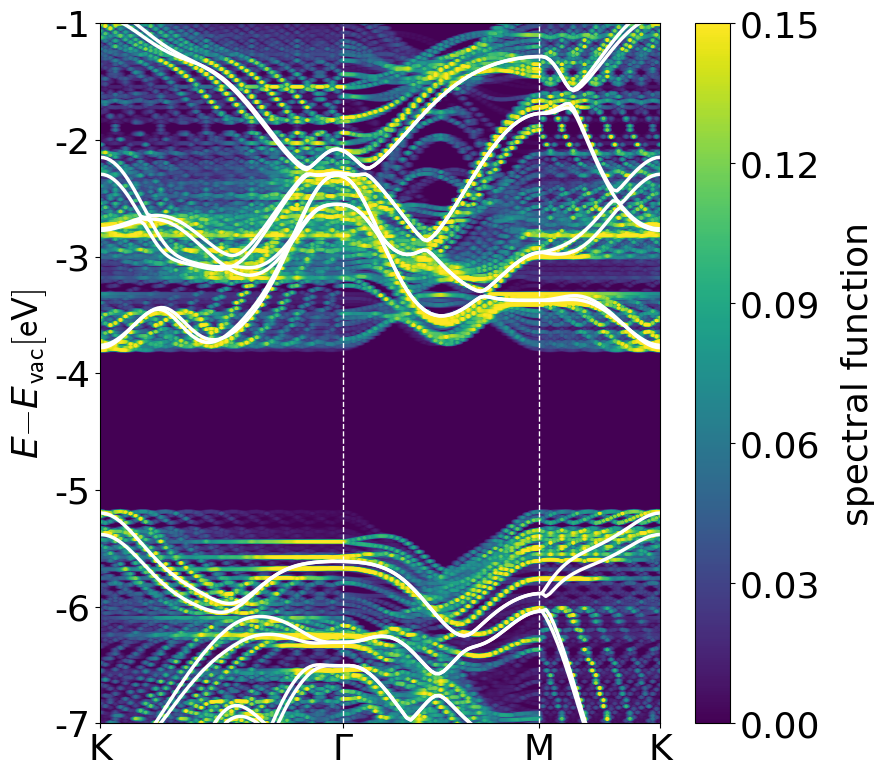}
\caption{Unfolded band structure of the nanowrinkle with $s=10\%$ represented by the spectral function and the corresponding values in the color scale. The white lines indicate the band structure of the flat and unstrained \ce{MoSe2} monolayer. The energy scale is offset with respect to the vacuum level ($E_{\textrm{vac}}$) set to zero.
}
\label{fig:bs}
\end{figure}

In the next step of our analysis, we inspect the band structures (BS) of the nanowrinkles to understand the impact of local and global strain, including curvature, on their electronic properties.
It is convenient to unfold these results onto a hexagonal representation to enable a direct comparison with the band structure of pristine \ce{MoSe2} as well as with its strained but flat counterparts, and thus to single out the curvature effects in the electronic properties of the nanowrinkles. 
In Fig.~\ref{fig:bs}, we show the unfolded band structure of the nanowrinkle with $s = 10\%$ (results obtained for the other systems are reported in Figs.~S4 and S5).
In the supercell, the translational symmetry of the original UC is broken. Consequently, the eigenstates that were originally symmetric (up to a phase factor) under the UC translation operator are now split into multiple components with different k-vectors. In this representation, each of such components is displayed by the value of its spectral function defined in Eq.~\eqref{eq:spectral}. It should be stressed that the unfolded band structure does not provide a clear insight into the (in)direct nature of the bandgap. In this picture, optical transitions are not required to be vertical in k-space, as a consequence of the reduced translational symmetry compared to the orthogonal representation adopted for calculating the nanowrinkles.
In Fig.~\ref{fig:bs}, both the valence-band maximum (VBM) and the conduction-band minimum (CBm) are delocalized in k-space. However, the largest spectral weights are found at the high-symmetry point K, which coincides with the VBM and CBm also in the flat, unstrained monolayer in the hexagonal representation (see Figs.~S6 and S7). 
In the unfolded band structure of the nanowrinkle, the VBM and CBm are accompanied by a number of replicas with lower spectral weight due to the loss of the translation symmetry of the pristine unit cell (see Fig.~S2).
In the unoccupied region, differences in the spectral weight of the lowest bands at $\Gamma$ and K, where the CBm of pristine \ce{MoSe2} is located, are less pronounced.

The size of the fundamental gap of \ce{MoSe2} does not seem to be significantly affected in the nanowrinkle (1.39~eV) with $s = 10\%$ compared to the pristine monolayer, where it amounts to 1.36~eV~\cite{isla+16ns}.
Likewise, its position remains at the high-symmetry point K, see Fig.~\ref{fig:bs}.
In the valence region, the band dispersion between K and $\Gamma$ of the uppermost, SOC-split band that is visible in the BS of the unstrained sheet (white lines in Fig.~\ref{fig:bs}) is present also in the nanowrinkle.
In addition, in the latter system, a few flat bands appear along the K-$\Gamma$ path between $-5.4$~eV and $-5.75$~eV; their largest spectral weight appears in the vicinity of the $\Gamma$ point.
These states are related to the highest valence band of the pristine monolayer which has a region of low dispersion around the BZ center. 
Between $\Gamma$ and M, the nanowrinkle, as well as the flat and unstrained monolayer, are characterized by dispersive states: in the former, they appear at higher energies compared to the latter.
A major difference in the electronic structure of these two systems appears between M and K.
In the flat monolayer, a SOC-split band with a large, positive slope, spans the interval $-5.90$--$-5.20$~eV.
In the nanowrinkle, in the same energy range, several flat bands are visible in Fig.~\ref{fig:bs}.
At lower energies in the valence region, the bands of unstrained TMDC are again largely replicated in the BS of the nanowrinkle, similarly as in the energy range above.

In the conduction region, the highly dispersive state of pristine \ce{MoSe2} that is present at the band edge is replaced by a plethora of superimposed bands in the rippled system.
The K-$\Gamma$ as well as the M-K paths are dominated by numerous replica-like components of the same state, all lying at the same energy as the CBm.
Only between $\Gamma$ and M is the minimum in the BS of the flat monolayer replicated in the nanowrinkle about 240~meV lower in energy, see Fig.~\ref{fig:bs}.
The high-energy states in the conduction band follow the same pattern: the highly dispersive bands of the flat monolayer are replicated partly with slight energy shifts and partly by a large density of flat bands in the electronic structure of the nanowrinkle.

The peculiar characteristics of the band structure of the nanowrinkle with $s=10\%$ call for a deeper understanding of their origin.
Existing literature on strained but flat TMDCs reveals significant variations in the band gaps. For \ce{MoSe2} with 10\% biaxial strain, the gap has been noticed to decrease to 0~eV (tensile strain) ~\cite{joha-shen12nano}, 0.4~eV and 0.2~eV (tensile and compressive strain, respectively)~\cite{horz+13prb}, and 0.15~eV (tensile strain, extrapolated from the data in Ref.~\citenum{mani+16ssc}). In \ce{MoS2}, a semiconductor-to-metal transition was observed~\cite{scal+12nr} under both compressive (15\%) and tensile (10\%) strain. 
In contrast, the results for the nanowrinkle shown in Fig.~\ref{fig:bs} do not point in the direction of such drastic changes but rather toward a dense distribution of the electronic states in the vicinity of the gap, which decreases by only a few tens of meV compared to the pristine material, as pointed out above. 
Similar results, albeit in a slightly different setup, are consistent with those presented in Refs.~\cite{nepal2019first,neupane2021opening}.
Based on this finding, it is legitimate to speculate that the present constraint of the local strain to small regions of the nanowrinkle is responsible for the marginal modification of the gap despite the large value of strain applied.

\begin{figure}
    \centering
    \includegraphics[width=0.5\textwidth-9pt]{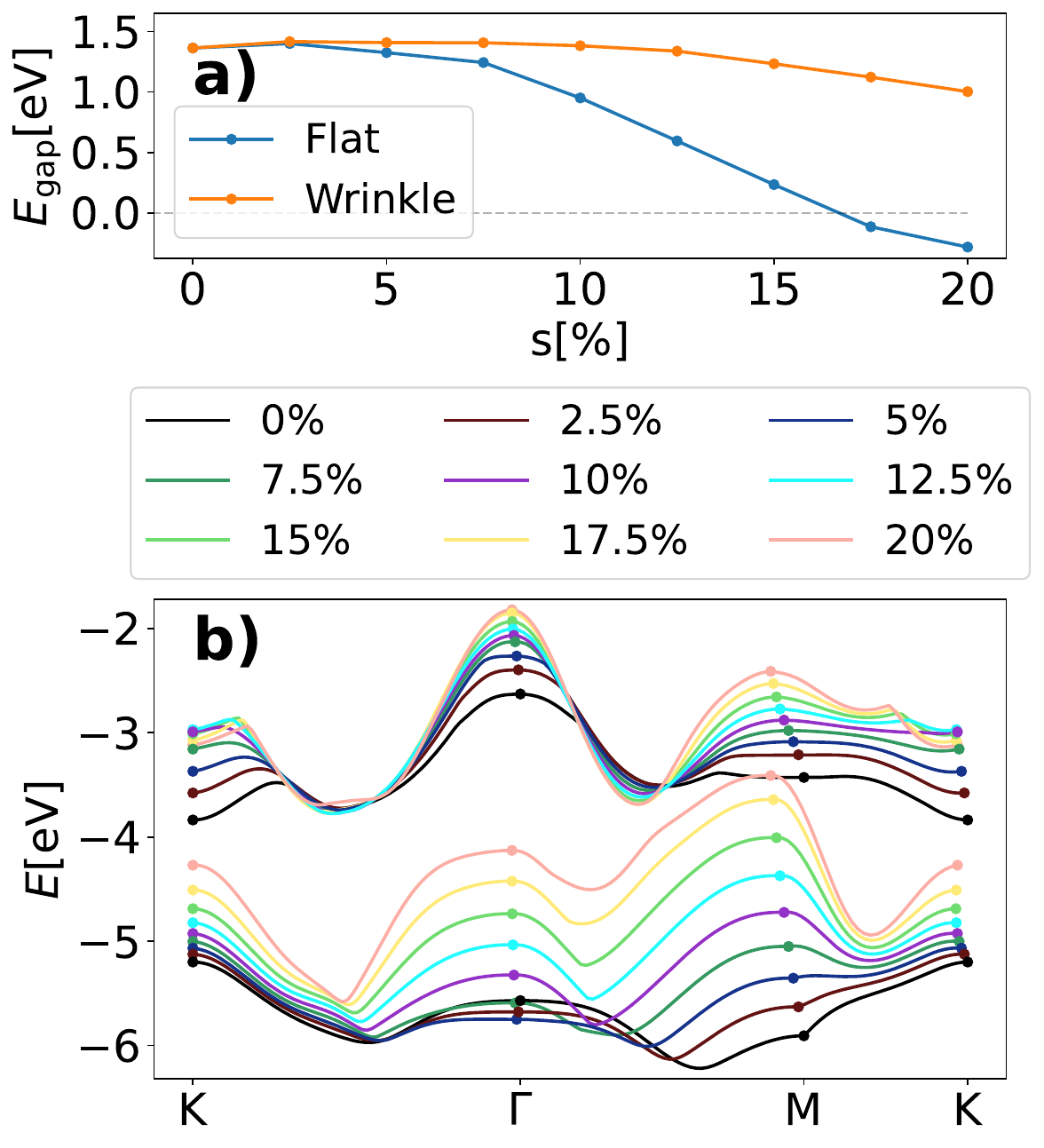}
    \caption{a) Band gap ($E_\textrm{gap}$) \textit{vs} strain ($s$) of the considered \ce{MoSe2} nanowrinkles (orange) and of the flat monolayers under the same amount of uniaxial strain in their hexagonal UC (blue). b) Highest-occupied and lowest-unoccupied band of flat \ce{MoSe2} under the displayed values of strain. The deformation of the BZ leads to different distances between the high-symmetry points in the K-path and hence they are shown separately for each strain value.}
    \label{fig:flatgap}
\end{figure}

To check this hypothesis, we examine the size of the band gap as a function of the strain $s$ in the considered configurations.
The results shown in Fig.~\ref{fig:flatgap}a) indicate a variation of 430~meV between the largest value ($E_{gap}=$1.46~eV) obtained for the system with $s=5\%$ and the smallest one ($E_{gap}=$1.02~eV) pertaining to the configuration with the largest strain, $s=20\%$.
It is worth noting that the trend is non-monotonic, especially for low values of $s$.
For $s>10\%$, the gap size decreases steadily with increasing strain. 

Next, we contrast these results with those obtained for flat \ce{MoSe2} sheets in their hexagonal UC subject to the same amounts of uniaxial strain.
In this case, the band-gap shrinks monotonically at increasing compressive deformation of the lattice, in agreement with previous findings~\cite{cheng2020using, deng+18pe}.
In particular, the system becomes a metal under uniaxial strain above 15\%.  
We recall that the values reported in Fig.~\ref{fig:flatgap}a) are obtained from DFT using the PBE functional, which is known to severely underestimate band gaps. However, this limitation uniformly manifests itself within similar systems (like the wrinkles examined in this paper), and as such, the derived trends are usually trustworthy. 
Under this disclaimer, the results presented and discussed in this section should be interpreted only on a qualitative level.

In Fig.~\ref{fig:flatgap}b), we visualize the highest valence band and the lowest conduction band of a strained \ce{MoSe2} monolayer in its flat geometry.
Our results indicate an upshift of both states.
However, this energy increase is particularly pronounced at the M-point of the valence band, such that already at moderate strain values ($s>2.5\%$), the gap becomes indirect and is found between M and K, in agreement with previous findings~\cite{horz+13prb}. Under extreme deformation, the system becomes metallic as the top of the valence band at M is energetically higher than the CBM between $\Gamma$ and M. 

\subsection{Correlation between effective mass and wave-function localization}
\label{ss:eff_mass}

\begin{figure}
\centering
\includegraphics[width=0.5\textwidth-9pt]{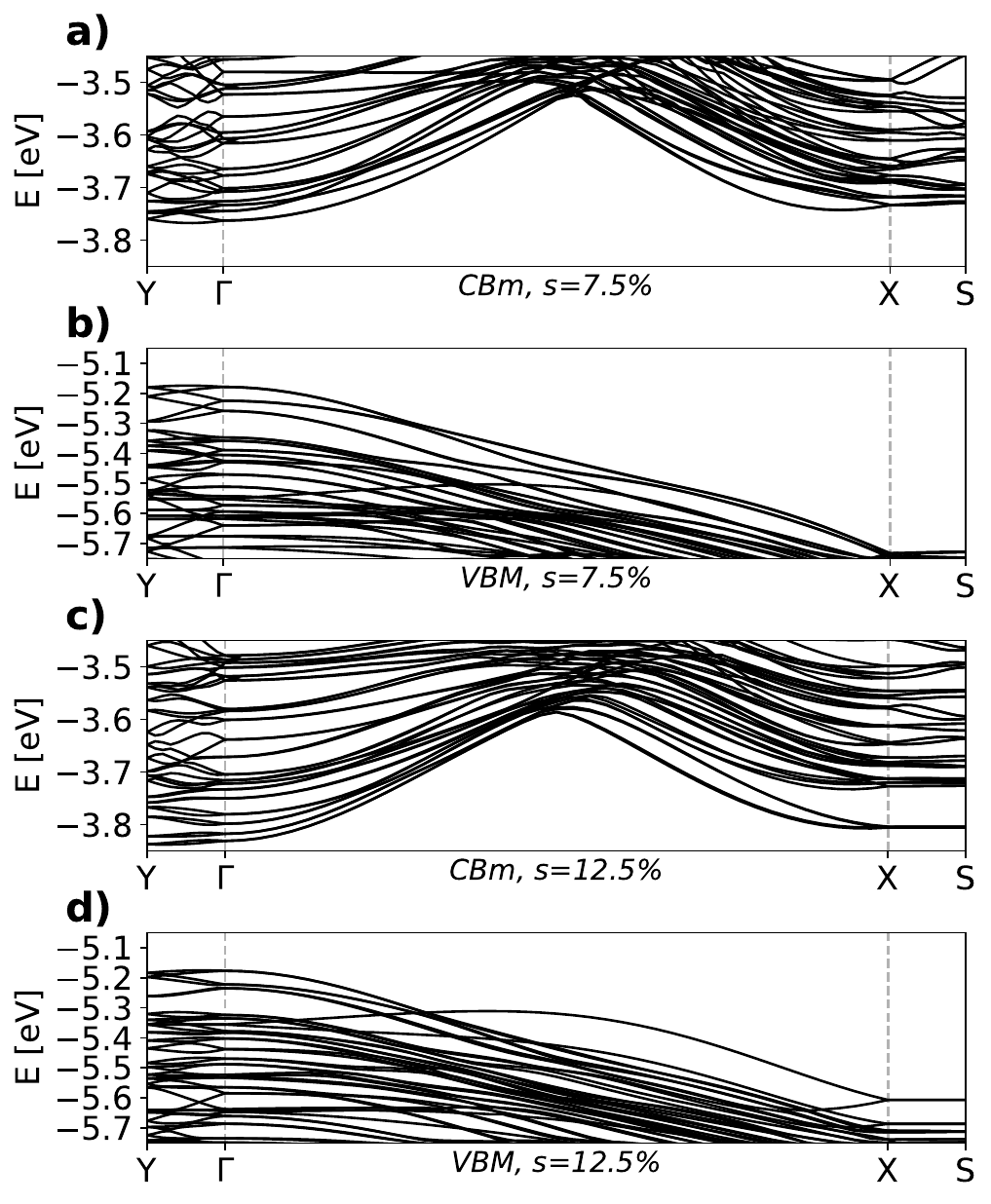}
\caption{Band structure in the orthorhombic BZ for the nanowrinkle with $s=7.5\%$ [conduction region in a) and valence region in b)] and $s=12.5\%$ [conduction region in c) and valence region in d)].} 
\label{fig:bs_folded}
\end{figure}

\begin{figure}
    \centering
    \includegraphics[width=0.45\textwidth]{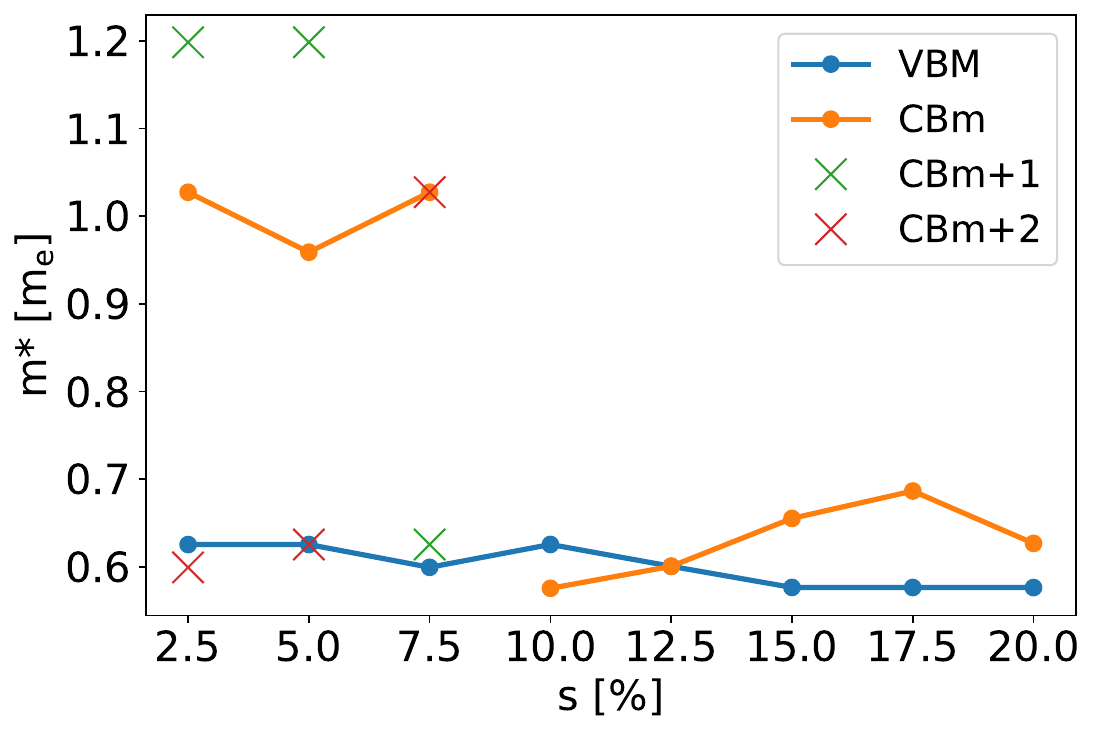}
    \caption{Effective masses of the highest valence band (VBM) and the lowest unoccupied states (CBm) at the $\Gamma$ point.}
    \label{fig:effmass_folded}
\end{figure}

The analysis of the electronic properties of the nanowrinkles in comparison with their flat counterparts reported above has clarified that deformation plays a vital role in preserving the semiconducting nature of \ce{MoSe2} even under amounts of strain above 10\%. 
The visualization of the band structure in the hexagonal UC facilitates the comparison with the pristine material but is not very practical for the examination of some of the intrinsic characteristics of the nanowrinkles.
In Fig.~\ref{fig:bs_folded}, we display the band structures of two selected systems with $s=7.5\%$ [Figs.~\ref{fig:bs_folded}a)-b)] and $s=12.5\%$ [Figs.~\ref{fig:effmass_folded}c)-d)] along the k-path of their native, orthorhombic BZ (see Fig.~S2). Corresponding results computed for all the systems under study are reported in the Supplementary Material, Fig.~S3).

From the inspection of these results, we immediately recognize the large number of states close to the gap, which we discussed extensively in Section~\ref{ss:bs}.
The VBM is at $\Gamma$, while the CBm appears at Y for $s=12.5\%$ [see Fig.~\ref{fig:bs_folded}c)].
The energy difference between the conduction band minima at $\Gamma$ and Y is, however, of the order of a few meV and does not affect the outcome of this discussion.
The flatness of the bands along the Y-$\Gamma$ can be rationalized by recalling that this path corresponds to the direction across the wrinkle in real space. The curves of the wrinkle act as structural defects in the lattice, which hamper charge carrier mobility.
In contrast, along the $\Gamma$-X path, valence and conduction bands in both considered systems exhibit large dispersion: in real space, this is the direction along the ridge of the nanowrinkle where we expect the largest charge mobility. 

Correlation between effective mass and mobility under varying biaxial strain has previously been examined computationally for \ce{MoS2}~\cite{yu2015phase}, which we expect to qualitatively match the properties of \ce{MoSe2}. Results from DFT are often used in conjunction with the Boltzmann equation or the Drude model to access transport properties of materials~\cite{mads-sing06cpc,transport-book}, for example, through the explicit evaluation of the current (density) or the conductivity which can be compared directly to experimental results. Due to the lack of measurements for the specific systems considered in this study, we focus the following analysis only on intrinsic material properties such as the effective mass. A connection to macroscopic models can be straightforwardly implemented as soon as experimental references become available. 

We evaluate the effective mass associated with the highest-occupied and lowest-unoccupied states at $\Gamma$, towards the X-point, which matches the direction along the ridge in real space. It should be noted that, while there is a twofold degeneracy of states at $\Gamma$, the degeneracy is lifted by SOC away from $\Gamma$, resulting in two different parabolic bands from which the effective masses are derived. In this analysis, we consider the lower of the paired states.
The results, displayed in Fig.~\ref{fig:effmass_folded}, indicate similar trends as a function of strain for electrons and holes.
The effective mass associated with the VBM at $\Gamma$ remains of the order of 0.6~$m_\mathrm{e}$ regardless of $s$. The same holds for the CBm, with minor variations and except for the first three values of $s$, where $m^*$ is of the order of 1.0~$m_\mathrm{e}$.
We propose that this sudden change in behavior at the CBm stems from an energetic reordering of states in the nanowrinkles with $s\geq 10\%$ and $s\leq 7.5\%$. 
This hypothesis is substantiated by examining, for $s\leq 7.5\%$, the $m^*$ values associated with the second and third lowest states in the conduction band at $\Gamma$, labeled in Fig.~\ref{fig:effmass_folded} as ``CBm+1" and ``CBm+2", respectively. The second lowest state for $s=7.5\%$ and the third lowest state for $s\leq 5\%$ indeed have the value of $m^*$ in the range of $~0.6m_\mathrm{e}$. These values of $m^* \approx 0.6 m_\mathrm{e}$ are in good agreement with existing literature for flat, unstrained \ce{MoSe2}~\cite{harada2014computational, yun2012thickness,chang2014ballistic}, showing that carrier mobility along the ridge of the nanowrinkle is expected to be similar to that of a pristine \ce{MoSe2} monolayer.
Further details are reported in the Supplementary Material, Fig.~S8.

\begin{figure}
    \centering
    \includegraphics[width=0.45\textwidth]{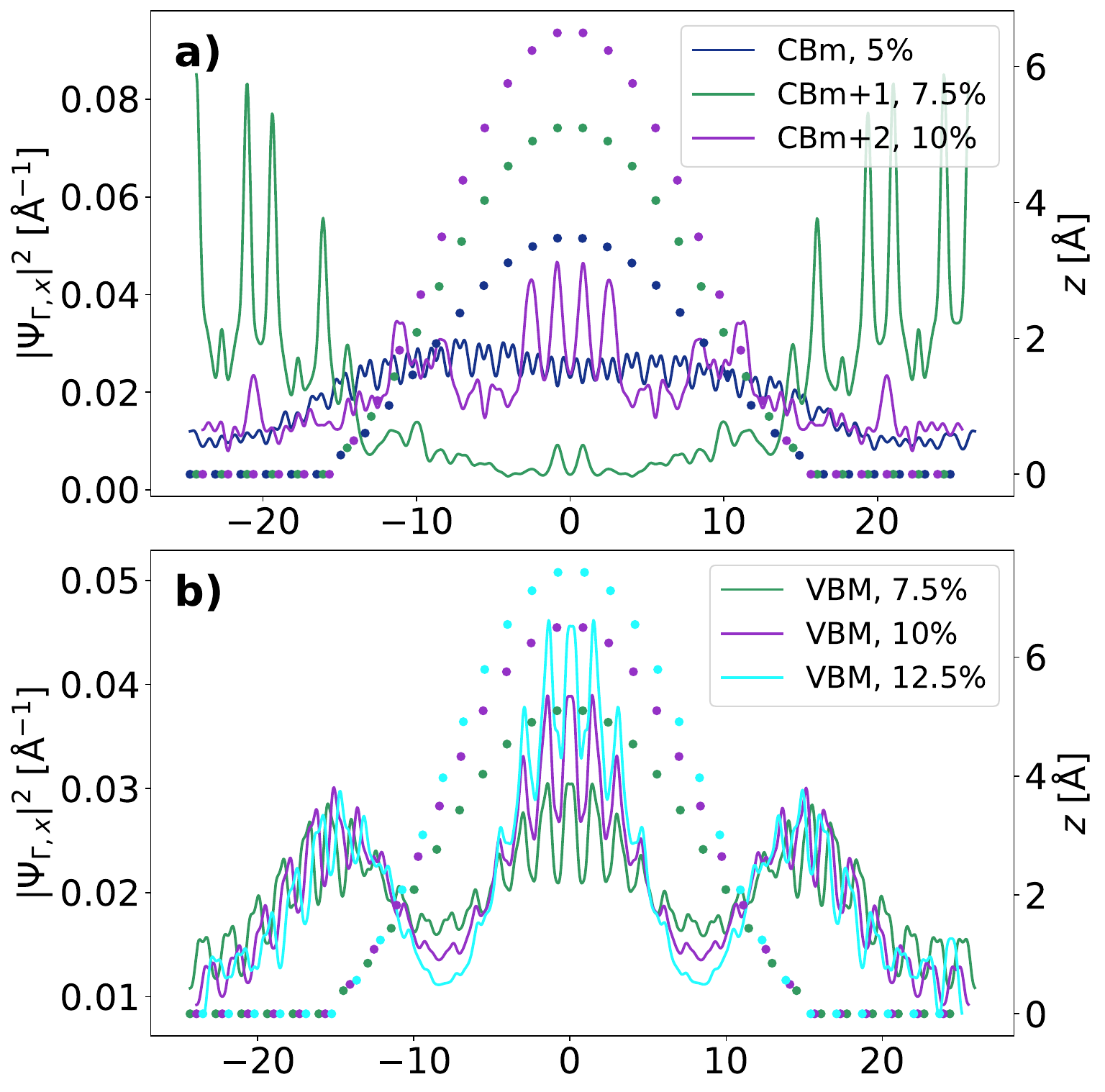}
    \caption{Normalized wave-function distribution averaged in the $(y,z)$-plane of the a) CBm and b) VBM at $\Gamma$ calculated for selected nanowrinkles with $s=7.5\%$, $s=10\%$, and $s=12.5\%$. The $z$-coordinates of the Mo atoms in the nanowrinkles are shown for reference by colored dots. }
    \label{fig:wfd}
\end{figure}

We deepen our analysis by inspecting the square modulus $|\Psi|^2$ of the wave-function, averaged in the ($y,z$)-plane, associated with the highest valence states and the three lowest conduction states at $\Gamma$. This averaged quantity will be referred to as the wave-function distribution (WFD) and denoted by $|\Psi_{\Gamma, x}|^\mathsf{2}$. 
In the valence region, the density of the highest state is rather homogeneously distributed along the $x$-direction for the system with $s=7.5\%$, Fig.~\ref{fig:wfd}b). In general, for values of $s$ up to $12.5\%$, electrons are more likely to be found in the high-curvature regions (peak and bases) of the nanowrinkle than in the rest of the structure. For higher strains ($s\geq 15\%$), electrons are almost exclusively localized at the peak (see Fig.~S9).

For the conduction band, we identify three types of WFD that are relevant for the analysis of the effective mass: delocalized, localized, and semi-localized. Corresponding examples are shown in Fig.~\ref{fig:wfd}a). It can be seen that the indigo-coloured WFD, corresponding to the CBm at $s=5\%$, is relatively evenly spread throughout the wrinkle and is considered delocalized. In contrast, the green-colored WFD, at the second lowest state at $s=7.5\%$, is localized on the periphery of the nanowrinkle. The purple-colored line, representing the third lowest state at $s=10\%$ shows some features resembling both the localized and delocalized WFD and is classified as semi-localized. The complete data for all the considered WFD are shown in Figs.~S10-S12.

\begin{figure}
    \centering
    \includegraphics[width=0.48\textwidth]{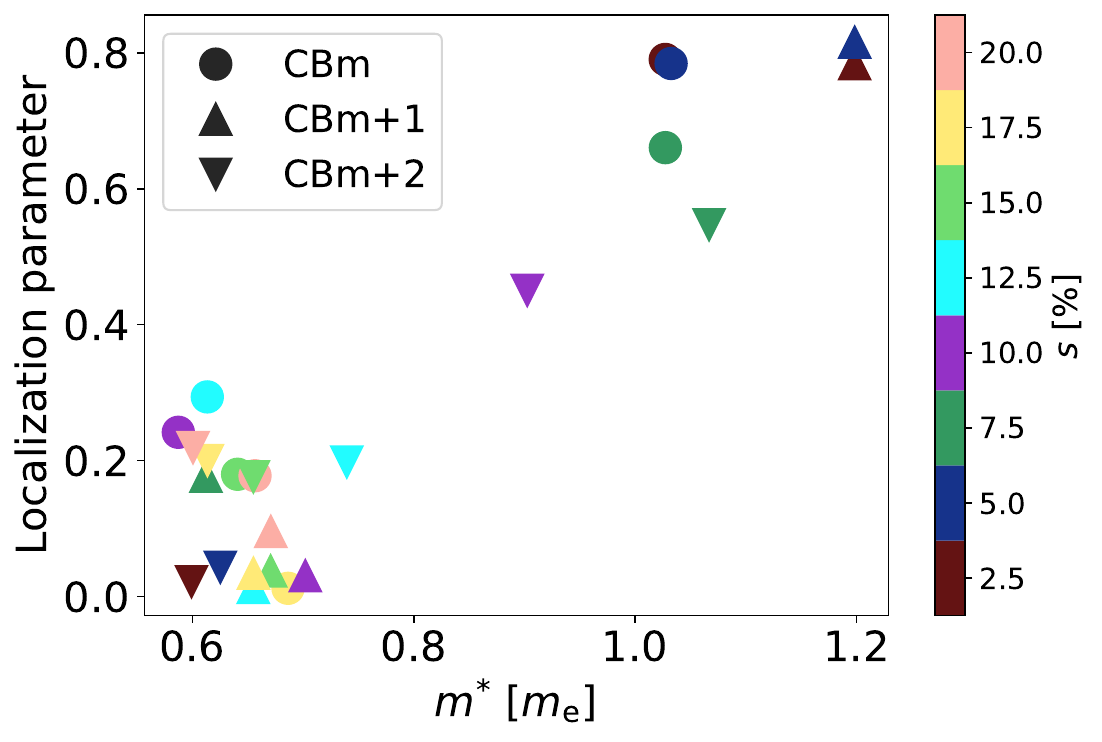}
    \caption{Localization parameter $L$ mapped against effective mass $m^*$ for the 3 lowest states in the conduction band at $\Gamma$. States with $m^* < 0.8 \mathrm{m}_\mathrm{e}$ are densely clustered in the area with low $L$, while the ones with higher effective mass are more dispersed, but still clearly correspond to higher values of $L$.}
    \label{fig:L_m_correlation}
\end{figure}

Based on the behavior of the WFD of those states, we identify a correlation between the localization of the wave-function distribution of the $\Gamma$ state and its corresponding effective mass. More precisely, we introduce a unitless localization parameter $L$, which provides a quantitative description of whether the wave-function distribution is localized, semi-localized, or delocalized by taking into account its spatial distribution. $L$ can approximately be thought of as the absolute difference between the normalized standard deviation of the WFD and a similarly normalized measure of the similarity of its points to their neighborhood. An explicit description of $L$ is given in the Supplementary Material (Eqs.~S1 and S2). A low value of $L$ indicates high localization and corresponds to states with low effective mass. The values of both above-mentioned quantities for the 3 lowest unoccupied states at $\Gamma$ and all strains are plotted in Fig.~\ref{fig:L_m_correlation}. The distribution of all those values of $m^*$ along $s$ is displayed in Fig.~S13.

\begin{figure}
    \centering
    \includegraphics[width=0.48\textwidth]{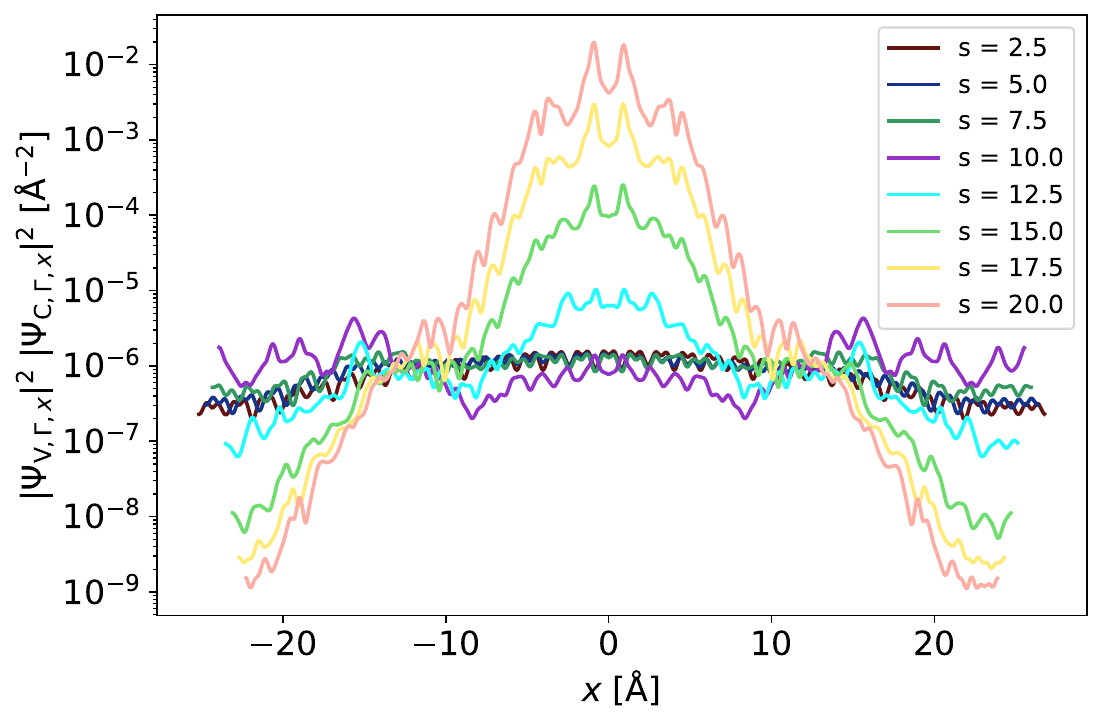}
    \caption{Product of the WFD values at the frontier states, plotted on a logarithmic scale. Strong localization at higher values of strain leads to a very large overlap of the wave functions. The V and C indices denote valence and conduction bands, respectively.}
    \label{fig:WFD_overlap}
\end{figure}

The results of this analysis have important implications. 
Under lower strain values of $s\leq 7.5\%$, the broad distribution of the highest valence state overlaps with the lowest conduction state throughout the entire simulation cell. For $s=10\%$, the peaked structure of the WFD associated with the VBM overlaps well with its CBm counterpart at the bases of the nanowrinkle. As the strain increases further, the localization of both frontier states at the peak becomes even stronger, resulting in a very pronounced overlap within a confined region. This overlap is represented as a product of the square moduli of the corresponding wave functions and is visualized in Fig.~\ref{fig:WFD_overlap}. 
We emphasize that these characteristics have implications on the intensity and spatial distribution of excitons generated in these systems. Future work beyond DFT is required to shed light on these aspects. However, the findings presented here can be already used to interpret experimental data.

\section{Summary and conclusions}
\label{sec:conclu}

In summary, we presented a detailed analysis of the electronic properties of \ce{MoSe2} nanowrinkles obtained by applying to the monolayer increasing values of strain between 2.5\% and 20\%.
The obtained structures with vertical elevations ranging from 0.1~\AA{} to 10.2~\AA{} and alternating local domains of compressive and tensile strain coexisting in the same materials when the strain $s$ exceeds 10\%. 
All the considered nanowrinkles are semiconductors and their direct band gaps are preserved in nature and magnitude despite the amount of deformation. Variations in magnitude are up to 94~meV for strains up to 10\% and up to 220~meV for $s\leq 15\%$. Only for the extreme strain values of 17.5\% and 20\% do the variations reach the values of 350~meV and 450~meV, respectively. 
A systematic comparison with flat \ce{MoSe2} sheets subject to the same amount of uniaxial strain as the nanowrinkles reveals that the curvature of the latter enables the conservation of the above-mentioned characteristics of the gap.
As expected, the flat, strained monolayers undergo a sizeable decrease of the band gap becoming metallic for extreme deformations above 15\%; furthermore, a direct-to-indirect band-gap transition occurs already for $s>2.5\%$.

We additionally calculated the effective mass for the frontier states at $\Gamma$ and found completely different behaviors in the valence and conduction region as a function of strain.
The effective mass for both electrons and holes remains of the order of 0.6~$m_\mathrm{e}$ for all considered nanowrinkles, except for those with $s\leq 7.5\%$, where the effective mass of the electrons is of the order of 1.0~$m_\mathrm{e}$. Based on the effective mass values at the second- and third-lowest conduction states at $\Gamma$, we speculate that this change of behavior is caused by the reordering of the bands at $s\leq 7.5\%$ compared to the systems with higher strain values. 
The analysis of the wave-function distribution of the frontier states provides additional insight. In particular, we managed to establish and quantify a correlation between the effective mass values and localization of the WFD at the three lowest conduction states.
Finally, we inspected the overlap between the electron and hole probability densities at the frontier states and expect high optical activity for $s\geq 12.5\%$ at the peak of the wrinkle, due to very strong charge carrier localization.

Our results offer a microscopic understanding of the electronic characteristics of TMDC nanowrinkles and of the structure-property relationships therein. 
Our findings provide indications to predict the optical response of these nanostructures, especially in the context of integrated optics and photonics, where they are currently pursued as active materials for single photon emission.
Intriguing perspectives emerge also for optoelectronic and spintronic applications~\cite{chou+19cm}, where the possibility to localize charge carrier on quasi-one-dimensional channels rather than on quantum-dot-like structures~\cite{krumland2024quantum} appears particularly promising.
While the systems simulated in this work are admittedly smaller than currently nanofabricated samples, the rationale obtained from our analysis can be scaled up to realistic systems as long as they are dominated by quantum-mechanical effects. 
Future work will be dedicated to obtaining quantitative estimates of relevant quantities such as fundamental and optical gaps using more advanced methods than DFT. 
Nonetheless, the identified connections between curvature, local strain domains, and electronic properties represent invaluable insight that is hardly achievable with any other methods.

\section*{Acknowledgments}
This work was funded by the QuanterERA II European Union’s Horizon 2020 research and innovation programme under the EQUAISE project, Grant Agreement No. 101017733, by the German Research Foundation (DFG), project number 182087777 -- CRC 951, by the German Federal Ministry of Education and Research (Professorinnenprogramm III), and by the State of Lower Saxony (Professorinnen für Niedersachsen). SV appreciates additional support from the Humboldt Internship Program in the early stage of this project. Computational resources were provided by the North-German Supercomputing Alliance (HLRN), project nip00063.

\section*{Data Availability}
The data that support the findings of this study are openly available in Zenodo at DOI 10.5281/zenodo.10260163. 



\begin{thebibliography}{78}%
\makeatletter
\providecommand \@ifxundefined [1]{%
 \@ifx{#1\undefined}
}%
\providecommand \@ifnum [1]{%
 \ifnum #1\expandafter \@firstoftwo
 \else \expandafter \@secondoftwo
 \fi
}%
\providecommand \@ifx [1]{%
 \ifx #1\expandafter \@firstoftwo
 \else \expandafter \@secondoftwo
 \fi
}%
\providecommand \natexlab [1]{#1}%
\providecommand \enquote  [1]{``#1''}%
\providecommand \bibnamefont  [1]{#1}%
\providecommand \bibfnamefont [1]{#1}%
\providecommand \citenamefont [1]{#1}%
\providecommand \href@noop [0]{\@secondoftwo}%
\providecommand \href [0]{\begingroup \@sanitize@url \@href}%
\providecommand \@href[1]{\@@startlink{#1}\@@href}%
\providecommand \@@href[1]{\endgroup#1\@@endlink}%
\providecommand \@sanitize@url [0]{\catcode `\\12\catcode `\$12\catcode
  `\&12\catcode `\#12\catcode `\^12\catcode `\_12\catcode `\%12\relax}%
\providecommand \@@startlink[1]{}%
\providecommand \@@endlink[0]{}%
\providecommand \url  [0]{\begingroup\@sanitize@url \@url }%
\providecommand \@url [1]{\endgroup\@href {#1}{\urlprefix }}%
\providecommand \urlprefix  [0]{URL }%
\providecommand \Eprint [0]{\href }%
\providecommand \doibase [0]{https://doi.org/}%
\providecommand \selectlanguage [0]{\@gobble}%
\providecommand \bibinfo  [0]{\@secondoftwo}%
\providecommand \bibfield  [0]{\@secondoftwo}%
\providecommand \translation [1]{[#1]}%
\providecommand \BibitemOpen [0]{}%
\providecommand \bibitemStop [0]{}%
\providecommand \bibitemNoStop [0]{.\EOS\space}%
\providecommand \EOS [0]{\spacefactor3000\relax}%
\providecommand \BibitemShut  [1]{\csname bibitem#1\endcsname}%
\let\auto@bib@innerbib\@empty
\bibitem [{\citenamefont {Shi}\ \emph {et~al.}(2015)\citenamefont {Shi},
  \citenamefont {Li},\ and\ \citenamefont {Li}}]{shi+15csr}%
  \BibitemOpen
  \bibfield  {author} {\bibinfo {author} {\bibfnamefont {Y.}~\bibnamefont
  {Shi}}, \bibinfo {author} {\bibfnamefont {H.}~\bibnamefont {Li}},\ and\
  \bibinfo {author} {\bibfnamefont {L.-J.}\ \bibnamefont {Li}},\ }\bibfield
  {title} {\bibinfo {title} {Recent advances in controlled synthesis of
  two-dimensional transition metal dichalcogenides via vapour deposition
  techniques},\ }\href@noop {} {\bibfield  {journal} {\bibinfo  {journal}
  {Chem.~Soc.~Rev.}\ }\textbf {\bibinfo {volume} {44}},\ \bibinfo {pages}
  {2744} (\bibinfo {year} {2015})}\BibitemShut {NoStop}%
\bibitem [{\citenamefont {Brar}\ \emph {et~al.}(2018)\citenamefont {Brar},
  \citenamefont {Sherrott},\ and\ \citenamefont {Jariwala}}]{brar+18csr}%
  \BibitemOpen
  \bibfield  {author} {\bibinfo {author} {\bibfnamefont {V.~W.}\ \bibnamefont
  {Brar}}, \bibinfo {author} {\bibfnamefont {M.~C.}\ \bibnamefont {Sherrott}},\
  and\ \bibinfo {author} {\bibfnamefont {D.}~\bibnamefont {Jariwala}},\
  }\bibfield  {title} {\bibinfo {title} {Emerging photonic architectures in
  two-dimensional opto-electronics},\ }\href@noop {} {\bibfield  {journal}
  {\bibinfo  {journal} {Chem.~Soc.~Rev.}\ }\textbf {\bibinfo {volume} {47}},\
  \bibinfo {pages} {6824} (\bibinfo {year} {2018})}\BibitemShut {NoStop}%
\bibitem [{\citenamefont {Ko}\ \emph {et~al.}(2020)\citenamefont {Ko},
  \citenamefont {Wang}, \citenamefont {Yoo}, \citenamefont {Okogbue},
  \citenamefont {Islam}, \citenamefont {Li}, \citenamefont {Shawkat},
  \citenamefont {Han}, \citenamefont {Oh},\ and\ \citenamefont
  {Jung}}]{ko+20jpd}%
  \BibitemOpen
  \bibfield  {author} {\bibinfo {author} {\bibfnamefont {T.-J.}\ \bibnamefont
  {Ko}}, \bibinfo {author} {\bibfnamefont {M.}~\bibnamefont {Wang}}, \bibinfo
  {author} {\bibfnamefont {C.}~\bibnamefont {Yoo}}, \bibinfo {author}
  {\bibfnamefont {E.}~\bibnamefont {Okogbue}}, \bibinfo {author} {\bibfnamefont
  {M.~A.}\ \bibnamefont {Islam}}, \bibinfo {author} {\bibfnamefont
  {H.}~\bibnamefont {Li}}, \bibinfo {author} {\bibfnamefont {M.~S.}\
  \bibnamefont {Shawkat}}, \bibinfo {author} {\bibfnamefont {S.~S.}\
  \bibnamefont {Han}}, \bibinfo {author} {\bibfnamefont {K.~H.}\ \bibnamefont
  {Oh}},\ and\ \bibinfo {author} {\bibfnamefont {Y.}~\bibnamefont {Jung}},\
  }\bibfield  {title} {\bibinfo {title} {Large-area 2{D} {TMD} layers for
  mechanically reconfigurable electronic devices},\ }\href@noop {} {\bibfield
  {journal} {\bibinfo  {journal} {J.~Phys.~D}\ }\textbf {\bibinfo {volume}
  {53}},\ \bibinfo {pages} {313002} (\bibinfo {year} {2020})}\BibitemShut
  {NoStop}%
\bibitem [{\citenamefont {Maiti}\ \emph {et~al.}(2020)\citenamefont {Maiti},
  \citenamefont {Patil}, \citenamefont {Saadi}, \citenamefont {Xie},
  \citenamefont {Azadani}, \citenamefont {Uluutku}, \citenamefont {Amin},
  \citenamefont {Briggs}, \citenamefont {Miscuglio}, \citenamefont
  {Van~Thourhout} \emph {et~al.}}]{mait+20natp}%
  \BibitemOpen
  \bibfield  {author} {\bibinfo {author} {\bibfnamefont {R.}~\bibnamefont
  {Maiti}}, \bibinfo {author} {\bibfnamefont {C.}~\bibnamefont {Patil}},
  \bibinfo {author} {\bibfnamefont {M.}~\bibnamefont {Saadi}}, \bibinfo
  {author} {\bibfnamefont {T.}~\bibnamefont {Xie}}, \bibinfo {author}
  {\bibfnamefont {J.}~\bibnamefont {Azadani}}, \bibinfo {author} {\bibfnamefont
  {B.}~\bibnamefont {Uluutku}}, \bibinfo {author} {\bibfnamefont
  {R.}~\bibnamefont {Amin}}, \bibinfo {author} {\bibfnamefont {A.}~\bibnamefont
  {Briggs}}, \bibinfo {author} {\bibfnamefont {M.}~\bibnamefont {Miscuglio}},
  \bibinfo {author} {\bibfnamefont {D.}~\bibnamefont {Van~Thourhout}}, \emph
  {et~al.},\ }\bibfield  {title} {\bibinfo {title} {Strain-engineered
  high-responsivity \ce{MoTe2} photodetector for silicon photonic integrated
  circuits},\ }\href@noop {} {\bibfield  {journal} {\bibinfo  {journal}
  {Nature~Phys.}\ }\textbf {\bibinfo {volume} {14}},\ \bibinfo {pages} {578}
  (\bibinfo {year} {2020})}\BibitemShut {NoStop}%
\bibitem [{\citenamefont {Gholipour}\ \emph {et~al.}(2023)\citenamefont
  {Gholipour}, \citenamefont {Elliott}, \citenamefont {M{\"u}ller},
  \citenamefont {Wuttig}, \citenamefont {Hewak}, \citenamefont {Hayden},
  \citenamefont {Li}, \citenamefont {Jo}, \citenamefont {Jaramillo},
  \citenamefont {Simpson} \emph {et~al.}}]{ghol+23jpp}%
  \BibitemOpen
  \bibfield  {author} {\bibinfo {author} {\bibfnamefont {B.}~\bibnamefont
  {Gholipour}}, \bibinfo {author} {\bibfnamefont {S.~R.}\ \bibnamefont
  {Elliott}}, \bibinfo {author} {\bibfnamefont {M.~J.}\ \bibnamefont
  {M{\"u}ller}}, \bibinfo {author} {\bibfnamefont {M.}~\bibnamefont {Wuttig}},
  \bibinfo {author} {\bibfnamefont {D.~W.}\ \bibnamefont {Hewak}}, \bibinfo
  {author} {\bibfnamefont {B.~E.}\ \bibnamefont {Hayden}}, \bibinfo {author}
  {\bibfnamefont {Y.}~\bibnamefont {Li}}, \bibinfo {author} {\bibfnamefont
  {S.~S.}\ \bibnamefont {Jo}}, \bibinfo {author} {\bibfnamefont
  {R.}~\bibnamefont {Jaramillo}}, \bibinfo {author} {\bibfnamefont {R.~E.}\
  \bibnamefont {Simpson}}, \emph {et~al.},\ }\bibfield  {title} {\bibinfo
  {title} {Roadmap on chalcogenide photonics},\ }\href@noop {} {\bibfield
  {journal} {\bibinfo  {journal} {J.~Phys.:~Photonics}\ }\textbf {\bibinfo
  {volume} {5}},\ \bibinfo {pages} {012501} (\bibinfo {year}
  {2023})}\BibitemShut {NoStop}%
\bibitem [{\citenamefont {Jeong}\ \emph {et~al.}(2017)\citenamefont {Jeong},
  \citenamefont {Oh}, \citenamefont {Gokarna}, \citenamefont {Kim},
  \citenamefont {Yun}, \citenamefont {Han}, \citenamefont {Jeong},
  \citenamefont {Lee},\ and\ \citenamefont {Lerondel}}]{jeon+17am}%
  \BibitemOpen
  \bibfield  {author} {\bibinfo {author} {\bibfnamefont {H.}~\bibnamefont
  {Jeong}}, \bibinfo {author} {\bibfnamefont {H.~M.}\ \bibnamefont {Oh}},
  \bibinfo {author} {\bibfnamefont {A.}~\bibnamefont {Gokarna}}, \bibinfo
  {author} {\bibfnamefont {H.}~\bibnamefont {Kim}}, \bibinfo {author}
  {\bibfnamefont {S.~J.}\ \bibnamefont {Yun}}, \bibinfo {author} {\bibfnamefont
  {G.~H.}\ \bibnamefont {Han}}, \bibinfo {author} {\bibfnamefont {M.~S.}\
  \bibnamefont {Jeong}}, \bibinfo {author} {\bibfnamefont {Y.~H.}\ \bibnamefont
  {Lee}},\ and\ \bibinfo {author} {\bibfnamefont {G.}~\bibnamefont
  {Lerondel}},\ }\bibfield  {title} {\bibinfo {title} {Integrated freestanding
  two-dimensional transition metal dichalcogenides},\ }\href@noop {} {\bibfield
   {journal} {\bibinfo  {journal} {Adv.~Mater.}\ }\textbf {\bibinfo {volume}
  {29}},\ \bibinfo {pages} {1700308} (\bibinfo {year} {2017})}\BibitemShut
  {NoStop}%
\bibitem [{\citenamefont {Parto}\ \emph {et~al.}(2021)\citenamefont {Parto},
  \citenamefont {Azzam}, \citenamefont {Banerjee},\ and\ \citenamefont
  {Moody}}]{part+21natcom}%
  \BibitemOpen
  \bibfield  {author} {\bibinfo {author} {\bibfnamefont {K.}~\bibnamefont
  {Parto}}, \bibinfo {author} {\bibfnamefont {S.~I.}\ \bibnamefont {Azzam}},
  \bibinfo {author} {\bibfnamefont {K.}~\bibnamefont {Banerjee}},\ and\
  \bibinfo {author} {\bibfnamefont {G.}~\bibnamefont {Moody}},\ }\bibfield
  {title} {\bibinfo {title} {Defect and strain engineering of monolayer
  \ce{WSe2} enables site-controlled single-photon emission up to 150 {K}},\
  }\href@noop {} {\bibfield  {journal} {\bibinfo  {journal} {Nature~Commun.}\
  }\textbf {\bibinfo {volume} {12}},\ \bibinfo {pages} {3585} (\bibinfo {year}
  {2021})}\BibitemShut {NoStop}%
\bibitem [{\citenamefont {Li}\ \emph {et~al.}(2022)\citenamefont {Li},
  \citenamefont {Chui}, \citenamefont {Shen}, \citenamefont {Huang},
  \citenamefont {Wen}, \citenamefont {Yam}, \citenamefont {Shao}, \citenamefont
  {Xu},\ and\ \citenamefont {Wang}}]{li+22nano}%
  \BibitemOpen
  \bibfield  {author} {\bibinfo {author} {\bibfnamefont {S.}~\bibnamefont
  {Li}}, \bibinfo {author} {\bibfnamefont {K.~K.}\ \bibnamefont {Chui}},
  \bibinfo {author} {\bibfnamefont {F.}~\bibnamefont {Shen}}, \bibinfo {author}
  {\bibfnamefont {H.}~\bibnamefont {Huang}}, \bibinfo {author} {\bibfnamefont
  {S.}~\bibnamefont {Wen}}, \bibinfo {author} {\bibfnamefont {C.}~\bibnamefont
  {Yam}}, \bibinfo {author} {\bibfnamefont {L.}~\bibnamefont {Shao}}, \bibinfo
  {author} {\bibfnamefont {J.}~\bibnamefont {Xu}},\ and\ \bibinfo {author}
  {\bibfnamefont {J.}~\bibnamefont {Wang}},\ }\bibfield  {title} {\bibinfo
  {title} {Generation and detection of strain-localized excitons in \ce{WS2}
  monolayer by plasmonic metal nanocrystals},\ }\href@noop {} {\bibfield
  {journal} {\bibinfo  {journal} {ACS~Nano}\ }\textbf {\bibinfo {volume}
  {16}},\ \bibinfo {pages} {10647} (\bibinfo {year} {2022})}\BibitemShut
  {NoStop}%
\bibitem [{\citenamefont {Kayal}\ \emph {et~al.}(2023)\citenamefont {Kayal},
  \citenamefont {Dey}, \citenamefont {G.}, \citenamefont {Nadarajan},
  \citenamefont {Chattopadhyay},\ and\ \citenamefont {Mitra}}]{kaya+23nl}%
  \BibitemOpen
  \bibfield  {author} {\bibinfo {author} {\bibfnamefont {A.}~\bibnamefont
  {Kayal}}, \bibinfo {author} {\bibfnamefont {S.}~\bibnamefont {Dey}}, \bibinfo
  {author} {\bibfnamefont {H.}~\bibnamefont {G.}}, \bibinfo {author}
  {\bibfnamefont {R.}~\bibnamefont {Nadarajan}}, \bibinfo {author}
  {\bibfnamefont {S.}~\bibnamefont {Chattopadhyay}},\ and\ \bibinfo {author}
  {\bibfnamefont {J.}~\bibnamefont {Mitra}},\ }\bibfield  {title} {\bibinfo
  {title} {Mobility enhancement in {CVD}-grown monolayer \ce{MoS2} via
  patterned substrate-induced nonuniform straining},\ }\href@noop {} {\bibfield
   {journal} {\bibinfo  {journal} {Nano~Lett.~}\ }\textbf {\bibinfo {volume}
  {23}},\ \bibinfo {pages} {6629} (\bibinfo {year} {2023})}\BibitemShut
  {NoStop}%
\bibitem [{\citenamefont {Blundo}\ \emph {et~al.}(2020)\citenamefont {Blundo},
  \citenamefont {Felici}, \citenamefont {Yildirim}, \citenamefont {Pettinari},
  \citenamefont {Tedeschi}, \citenamefont {Miriametro}, \citenamefont {Liu},
  \citenamefont {Ma}, \citenamefont {Lu},\ and\ \citenamefont
  {Polimeni}}]{blundo2020evidence}%
  \BibitemOpen
  \bibfield  {author} {\bibinfo {author} {\bibfnamefont {E.}~\bibnamefont
  {Blundo}}, \bibinfo {author} {\bibfnamefont {M.}~\bibnamefont {Felici}},
  \bibinfo {author} {\bibfnamefont {T.}~\bibnamefont {Yildirim}}, \bibinfo
  {author} {\bibfnamefont {G.}~\bibnamefont {Pettinari}}, \bibinfo {author}
  {\bibfnamefont {D.}~\bibnamefont {Tedeschi}}, \bibinfo {author}
  {\bibfnamefont {A.}~\bibnamefont {Miriametro}}, \bibinfo {author}
  {\bibfnamefont {B.}~\bibnamefont {Liu}}, \bibinfo {author} {\bibfnamefont
  {W.}~\bibnamefont {Ma}}, \bibinfo {author} {\bibfnamefont {Y.}~\bibnamefont
  {Lu}},\ and\ \bibinfo {author} {\bibfnamefont {A.}~\bibnamefont {Polimeni}},\
  }\bibfield  {title} {\bibinfo {title} {Evidence of the direct-to-indirect
  band gap transition in strained two-dimensional \ce{WS2}, \ce{MoS2}, and
  \ce{WSe2}},\ }\href@noop {} {\bibfield  {journal} {\bibinfo  {journal}
  {Phys.~Rev.~Res.}\ }\textbf {\bibinfo {volume} {2}},\ \bibinfo {pages}
  {012024} (\bibinfo {year} {2020})}\BibitemShut {NoStop}%
\bibitem [{\citenamefont {Zhang}\ \emph {et~al.}(2021)\citenamefont {Zhang},
  \citenamefont {Zhang}, \citenamefont {Qiu}, \citenamefont {Zhao},\ and\
  \citenamefont {Liu}}]{zhan+21im}%
  \BibitemOpen
  \bibfield  {author} {\bibinfo {author} {\bibfnamefont {W.}~\bibnamefont
  {Zhang}}, \bibinfo {author} {\bibfnamefont {Y.}~\bibnamefont {Zhang}},
  \bibinfo {author} {\bibfnamefont {J.}~\bibnamefont {Qiu}}, \bibinfo {author}
  {\bibfnamefont {Z.}~\bibnamefont {Zhao}},\ and\ \bibinfo {author}
  {\bibfnamefont {N.}~\bibnamefont {Liu}},\ }\bibfield  {title} {\bibinfo
  {title} {Topological structures of transition metal dichalcogenides: A review
  on fabrication, effects, applications, and potential},\ }\href@noop {}
  {\bibfield  {journal} {\bibinfo  {journal} {InfoMat}\ }\textbf {\bibinfo
  {volume} {3}},\ \bibinfo {pages} {133} (\bibinfo {year} {2021})}\BibitemShut
  {NoStop}%
\bibitem [{\citenamefont {Balgarkashi}\ \emph {et~al.}(2022)\citenamefont
  {Balgarkashi}, \citenamefont {Piazza}, \citenamefont {Jasi{\'n}ski},
  \citenamefont {Frisenda}, \citenamefont {Surrente}, \citenamefont
  {Baranowski}, \citenamefont {Dimitrievska}, \citenamefont {Dede},
  \citenamefont {Kim}, \citenamefont {G{\"u}niat} \emph
  {et~al.}}]{balg+22ieeejqe}%
  \BibitemOpen
  \bibfield  {author} {\bibinfo {author} {\bibfnamefont {A.}~\bibnamefont
  {Balgarkashi}}, \bibinfo {author} {\bibfnamefont {V.}~\bibnamefont {Piazza}},
  \bibinfo {author} {\bibfnamefont {J.}~\bibnamefont {Jasi{\'n}ski}}, \bibinfo
  {author} {\bibfnamefont {R.}~\bibnamefont {Frisenda}}, \bibinfo {author}
  {\bibfnamefont {A.}~\bibnamefont {Surrente}}, \bibinfo {author}
  {\bibfnamefont {M.}~\bibnamefont {Baranowski}}, \bibinfo {author}
  {\bibfnamefont {M.}~\bibnamefont {Dimitrievska}}, \bibinfo {author}
  {\bibfnamefont {D.}~\bibnamefont {Dede}}, \bibinfo {author} {\bibfnamefont
  {W.}~\bibnamefont {Kim}}, \bibinfo {author} {\bibfnamefont {L.}~\bibnamefont
  {G{\"u}niat}}, \emph {et~al.},\ }\bibfield  {title} {\bibinfo {title}
  {Spatial modulation of vibrational and luminescence properties of monolayer
  \ce{MoS2} using a gaas nanowire array},\ }\href@noop {} {\bibfield  {journal}
  {\bibinfo  {journal} {IEEE J. Quantum Electron.}\ }\textbf {\bibinfo {volume}
  {58}},\ \bibinfo {pages} {1} (\bibinfo {year} {2022})}\BibitemShut {NoStop}%
\bibitem [{\citenamefont {Shepard}\ \emph {et~al.}(2017)\citenamefont
  {Shepard}, \citenamefont {Ajayi}, \citenamefont {Li}, \citenamefont {Zhu},
  \citenamefont {Hone},\ and\ \citenamefont {Strauf}}]{shep+17_2DM}%
  \BibitemOpen
  \bibfield  {author} {\bibinfo {author} {\bibfnamefont {G.~D.}\ \bibnamefont
  {Shepard}}, \bibinfo {author} {\bibfnamefont {O.~A.}\ \bibnamefont {Ajayi}},
  \bibinfo {author} {\bibfnamefont {X.}~\bibnamefont {Li}}, \bibinfo {author}
  {\bibfnamefont {X.}~\bibnamefont {Zhu}}, \bibinfo {author} {\bibfnamefont
  {J.}~\bibnamefont {Hone}},\ and\ \bibinfo {author} {\bibfnamefont
  {S.}~\bibnamefont {Strauf}},\ }\bibfield  {title} {\bibinfo {title}
  {Nanobubble induced formation of quantum emitters in monolayer
  semiconductors},\ }\href@noop {} {\bibfield  {journal} {\bibinfo  {journal}
  {2{D} Mater.}\ }\textbf {\bibinfo {volume} {4}},\ \bibinfo {pages} {021019}
  (\bibinfo {year} {2017})}\BibitemShut {NoStop}%
\bibitem [{\citenamefont {Blundo}\ \emph {et~al.}(2021)\citenamefont {Blundo},
  \citenamefont {Yildirim}, \citenamefont {Pettinari},\ and\ \citenamefont
  {Polimeni}}]{blun+21prl}%
  \BibitemOpen
  \bibfield  {author} {\bibinfo {author} {\bibfnamefont {E.}~\bibnamefont
  {Blundo}}, \bibinfo {author} {\bibfnamefont {T.}~\bibnamefont {Yildirim}},
  \bibinfo {author} {\bibfnamefont {G.}~\bibnamefont {Pettinari}},\ and\
  \bibinfo {author} {\bibfnamefont {A.}~\bibnamefont {Polimeni}},\ }\bibfield
  {title} {\bibinfo {title} {Experimental adhesion energy in van der {W}aals
  crystals and heterostructures from atomically thin bubbles},\ }\href@noop {}
  {\bibfield  {journal} {\bibinfo  {journal} {Phys.~Rev.~Lett.}\ }\textbf
  {\bibinfo {volume} {127}},\ \bibinfo {pages} {046101} (\bibinfo {year}
  {2021})}\BibitemShut {NoStop}%
\bibitem [{\citenamefont {Kim}\ \emph {et~al.}(2022)\citenamefont {Kim},
  \citenamefont {Kim}, \citenamefont {Kumar}, \citenamefont {Rahaman},
  \citenamefont {Stevens}, \citenamefont {Jeon}, \citenamefont {Jo},
  \citenamefont {Kim}, \citenamefont {Trainor}, \citenamefont {Zhu} \emph
  {et~al.}}]{kim+22nano}%
  \BibitemOpen
  \bibfield  {author} {\bibinfo {author} {\bibfnamefont {G.}~\bibnamefont
  {Kim}}, \bibinfo {author} {\bibfnamefont {H.~M.}\ \bibnamefont {Kim}},
  \bibinfo {author} {\bibfnamefont {P.}~\bibnamefont {Kumar}}, \bibinfo
  {author} {\bibfnamefont {M.}~\bibnamefont {Rahaman}}, \bibinfo {author}
  {\bibfnamefont {C.~E.}\ \bibnamefont {Stevens}}, \bibinfo {author}
  {\bibfnamefont {J.}~\bibnamefont {Jeon}}, \bibinfo {author} {\bibfnamefont
  {K.}~\bibnamefont {Jo}}, \bibinfo {author} {\bibfnamefont {K.-H.}\
  \bibnamefont {Kim}}, \bibinfo {author} {\bibfnamefont {N.}~\bibnamefont
  {Trainor}}, \bibinfo {author} {\bibfnamefont {H.}~\bibnamefont {Zhu}}, \emph
  {et~al.},\ }\bibfield  {title} {\bibinfo {title} {High-density, localized
  quantum emitters in strained 2{D} semiconductors},\ }\href@noop {} {\bibfield
   {journal} {\bibinfo  {journal} {ACS~Nano}\ }\textbf {\bibinfo {volume}
  {16}},\ \bibinfo {pages} {9651} (\bibinfo {year} {2022})}\BibitemShut
  {NoStop}%
\bibitem [{\citenamefont {Dirnberger}\ \emph {et~al.}(2021)\citenamefont
  {Dirnberger}, \citenamefont {Ziegler}, \citenamefont {Faria~Junior},
  \citenamefont {Bushati}, \citenamefont {Taniguchi}, \citenamefont {Watanabe},
  \citenamefont {Fabian}, \citenamefont {Bougeard}, \citenamefont {Chernikov},\
  and\ \citenamefont {Menon}}]{dirn+21sa}%
  \BibitemOpen
  \bibfield  {author} {\bibinfo {author} {\bibfnamefont {F.}~\bibnamefont
  {Dirnberger}}, \bibinfo {author} {\bibfnamefont {J.~D.}\ \bibnamefont
  {Ziegler}}, \bibinfo {author} {\bibfnamefont {P.~E.}\ \bibnamefont
  {Faria~Junior}}, \bibinfo {author} {\bibfnamefont {R.}~\bibnamefont
  {Bushati}}, \bibinfo {author} {\bibfnamefont {T.}~\bibnamefont {Taniguchi}},
  \bibinfo {author} {\bibfnamefont {K.}~\bibnamefont {Watanabe}}, \bibinfo
  {author} {\bibfnamefont {J.}~\bibnamefont {Fabian}}, \bibinfo {author}
  {\bibfnamefont {D.}~\bibnamefont {Bougeard}}, \bibinfo {author}
  {\bibfnamefont {A.}~\bibnamefont {Chernikov}},\ and\ \bibinfo {author}
  {\bibfnamefont {V.~M.}\ \bibnamefont {Menon}},\ }\bibfield  {title} {\bibinfo
  {title} {Quasi-1{D} exciton channels in strain-engineered 2{D} materials},\
  }\href@noop {} {\bibfield  {journal} {\bibinfo  {journal} {Sci.~Adv.}\
  }\textbf {\bibinfo {volume} {7}},\ \bibinfo {pages} {eabj3066} (\bibinfo
  {year} {2021})}\BibitemShut {NoStop}%
\bibitem [{\citenamefont {Darlington}\ \emph {et~al.}(2020)\citenamefont
  {Darlington}, \citenamefont {Carmesin}, \citenamefont {Florian},
  \citenamefont {Yanev}, \citenamefont {Ajayi}, \citenamefont {Ardelean},
  \citenamefont {Rhodes}, \citenamefont {Ghiotto}, \citenamefont {Krayev},
  \citenamefont {Watanabe} \emph {et~al.}}]{darl+20natn}%
  \BibitemOpen
  \bibfield  {author} {\bibinfo {author} {\bibfnamefont {T.~P.}\ \bibnamefont
  {Darlington}}, \bibinfo {author} {\bibfnamefont {C.}~\bibnamefont
  {Carmesin}}, \bibinfo {author} {\bibfnamefont {M.}~\bibnamefont {Florian}},
  \bibinfo {author} {\bibfnamefont {E.}~\bibnamefont {Yanev}}, \bibinfo
  {author} {\bibfnamefont {O.}~\bibnamefont {Ajayi}}, \bibinfo {author}
  {\bibfnamefont {J.}~\bibnamefont {Ardelean}}, \bibinfo {author}
  {\bibfnamefont {D.~A.}\ \bibnamefont {Rhodes}}, \bibinfo {author}
  {\bibfnamefont {A.}~\bibnamefont {Ghiotto}}, \bibinfo {author} {\bibfnamefont
  {A.}~\bibnamefont {Krayev}}, \bibinfo {author} {\bibfnamefont
  {K.}~\bibnamefont {Watanabe}}, \emph {et~al.},\ }\bibfield  {title} {\bibinfo
  {title} {Imaging strain-localized excitons in nanoscale bubbles of monolayer
  \ce{WSe2} at room temperature},\ }\href@noop {} {\bibfield  {journal}
  {\bibinfo  {journal} {Nature~Nanotechnol.}\ }\textbf {\bibinfo {volume}
  {15}},\ \bibinfo {pages} {854} (\bibinfo {year} {2020})}\BibitemShut
  {NoStop}%
\bibitem [{\citenamefont {Long}\ \emph {et~al.}(2020)\citenamefont {Long},
  \citenamefont {Dai}, \citenamefont {Li},\ and\ \citenamefont
  {Jin}}]{long+20ns}%
  \BibitemOpen
  \bibfield  {author} {\bibinfo {author} {\bibfnamefont {C.}~\bibnamefont
  {Long}}, \bibinfo {author} {\bibfnamefont {Y.}~\bibnamefont {Dai}}, \bibinfo
  {author} {\bibfnamefont {J.}~\bibnamefont {Li}},\ and\ \bibinfo {author}
  {\bibfnamefont {H.}~\bibnamefont {Jin}},\ }\bibfield  {title} {\bibinfo
  {title} {Exciton manipulation in rippled transition metal dichalcogenides},\
  }\href@noop {} {\bibfield  {journal} {\bibinfo  {journal} {Nanoscale}\
  }\textbf {\bibinfo {volume} {12}},\ \bibinfo {pages} {21124} (\bibinfo {year}
  {2020})}\BibitemShut {NoStop}%
\bibitem [{\citenamefont {Jiang}\ and\ \citenamefont
  {Pachter}(2022)}]{jian-patc22ns}%
  \BibitemOpen
  \bibfield  {author} {\bibinfo {author} {\bibfnamefont {J.}~\bibnamefont
  {Jiang}}\ and\ \bibinfo {author} {\bibfnamefont {R.}~\bibnamefont
  {Pachter}},\ }\bibfield  {title} {\bibinfo {title} {Analysis of localized
  excitons in strained monolayer \ce{WSe2} by first principles calculations},\
  }\href@noop {} {\bibfield  {journal} {\bibinfo  {journal} {Nanoscale}\
  }\textbf {\bibinfo {volume} {14}},\ \bibinfo {pages} {11378} (\bibinfo {year}
  {2022})}\BibitemShut {NoStop}%
\bibitem [{\citenamefont {Lee}\ \emph {et~al.}(2023)\citenamefont {Lee},
  \citenamefont {Yun},\ and\ \citenamefont {Lee}}]{lee+23pccp}%
  \BibitemOpen
  \bibfield  {author} {\bibinfo {author} {\bibfnamefont {S.~Y.}\ \bibnamefont
  {Lee}}, \bibinfo {author} {\bibfnamefont {W.~S.}\ \bibnamefont {Yun}},\ and\
  \bibinfo {author} {\bibfnamefont {J.}~\bibnamefont {Lee}},\ }\bibfield
  {title} {\bibinfo {title} {Strain-induced dark exciton generation in rippled
  monolayer \ce{MoS2}},\ }\href@noop {} {\bibfield  {journal} {\bibinfo
  {journal} {Phys.~Chem.~Chem.~Phys.~}\ }\textbf {\bibinfo {volume} {25}},\
  \bibinfo {pages} {9894} (\bibinfo {year} {2023})}\BibitemShut {NoStop}%
\bibitem [{\citenamefont {Ghosh}\ \emph {et~al.}(2023)\citenamefont {Ghosh},
  \citenamefont {Papnai}, \citenamefont {Chen}, \citenamefont {Yadav},
  \citenamefont {Sankar}, \citenamefont {Hsieh}, \citenamefont {Hofmann},\ and\
  \citenamefont {Chen}}]{ghos+23am}%
  \BibitemOpen
  \bibfield  {author} {\bibinfo {author} {\bibfnamefont {R.}~\bibnamefont
  {Ghosh}}, \bibinfo {author} {\bibfnamefont {B.}~\bibnamefont {Papnai}},
  \bibinfo {author} {\bibfnamefont {Y.-S.}\ \bibnamefont {Chen}}, \bibinfo
  {author} {\bibfnamefont {K.}~\bibnamefont {Yadav}}, \bibinfo {author}
  {\bibfnamefont {R.}~\bibnamefont {Sankar}}, \bibinfo {author} {\bibfnamefont
  {Y.-P.}\ \bibnamefont {Hsieh}}, \bibinfo {author} {\bibfnamefont
  {M.}~\bibnamefont {Hofmann}},\ and\ \bibinfo {author} {\bibfnamefont {Y.-F.}\
  \bibnamefont {Chen}},\ }\bibfield  {title} {\bibinfo {title} {Exciton
  manipulation for enhancing photoelectrochemical hydrogen evolution reaction
  in wrinkled 2{D} heterostructures},\ }\href
  {https://doi.org/10.1002/adma.202210746} {\bibfield  {journal} {\bibinfo
  {journal} {Adv.~Mater.}\ ,\ \bibinfo {pages} {2210746}} (\bibinfo {year}
  {2023})}\BibitemShut {NoStop}%
\bibitem [{\citenamefont {Yu}\ \emph {et~al.}(2016)\citenamefont {Yu},
  \citenamefont {Ruzsinszky},\ and\ \citenamefont {Perdew}}]{yu+16nl}%
  \BibitemOpen
  \bibfield  {author} {\bibinfo {author} {\bibfnamefont {L.}~\bibnamefont
  {Yu}}, \bibinfo {author} {\bibfnamefont {A.}~\bibnamefont {Ruzsinszky}},\
  and\ \bibinfo {author} {\bibfnamefont {J.~P.}\ \bibnamefont {Perdew}},\
  }\bibfield  {title} {\bibinfo {title} {Bending two-dimensional materials to
  control charge localization and {F}ermi-level shift},\ }\href@noop {}
  {\bibfield  {journal} {\bibinfo  {journal} {Nano~Lett.~}\ }\textbf {\bibinfo
  {volume} {16}},\ \bibinfo {pages} {2444} (\bibinfo {year}
  {2016})}\BibitemShut {NoStop}%
\bibitem [{\citenamefont {Carmesin}\ \emph {et~al.}(2019)\citenamefont
  {Carmesin}, \citenamefont {Lorke}, \citenamefont {Florian}, \citenamefont
  {Erben}, \citenamefont {Schulz}, \citenamefont {Wehling},\ and\ \citenamefont
  {Jahnke}}]{carm+19nl}%
  \BibitemOpen
  \bibfield  {author} {\bibinfo {author} {\bibfnamefont {C.}~\bibnamefont
  {Carmesin}}, \bibinfo {author} {\bibfnamefont {M.}~\bibnamefont {Lorke}},
  \bibinfo {author} {\bibfnamefont {M.}~\bibnamefont {Florian}}, \bibinfo
  {author} {\bibfnamefont {D.}~\bibnamefont {Erben}}, \bibinfo {author}
  {\bibfnamefont {A.}~\bibnamefont {Schulz}}, \bibinfo {author} {\bibfnamefont
  {T.~O.}\ \bibnamefont {Wehling}},\ and\ \bibinfo {author} {\bibfnamefont
  {F.}~\bibnamefont {Jahnke}},\ }\bibfield  {title} {\bibinfo {title}
  {Quantum-dot-like states in molybdenum disulfide nanostructures due to the
  interplay of local surface wrinkling, strain, and dielectric confinement},\
  }\href@noop {} {\bibfield  {journal} {\bibinfo  {journal} {Nano~Lett.~}\
  }\textbf {\bibinfo {volume} {19}},\ \bibinfo {pages} {3182} (\bibinfo {year}
  {2019})}\BibitemShut {NoStop}%
\bibitem [{\citenamefont {Wang}\ \emph {et~al.}(2020)\citenamefont {Wang},
  \citenamefont {Ukhtary},\ and\ \citenamefont {Saito}}]{wang2020strain}%
  \BibitemOpen
  \bibfield  {author} {\bibinfo {author} {\bibfnamefont {S.}~\bibnamefont
  {Wang}}, \bibinfo {author} {\bibfnamefont {M.~S.}\ \bibnamefont {Ukhtary}},\
  and\ \bibinfo {author} {\bibfnamefont {R.}~\bibnamefont {Saito}},\ }\bibfield
   {title} {\bibinfo {title} {Strain effect on circularly polarized
  electroluminescence in transition metal dichalcogenides},\ }\href@noop {}
  {\bibfield  {journal} {\bibinfo  {journal} {Phys.~Rev.~Res.}\ }\textbf
  {\bibinfo {volume} {2}},\ \bibinfo {pages} {033340} (\bibinfo {year}
  {2020})}\BibitemShut {NoStop}%
\bibitem [{\citenamefont {Yu}\ \emph {et~al.}(2023)\citenamefont {Yu},
  \citenamefont {Dong}, \citenamefont {Binder}, \citenamefont {Schumacher},\
  and\ \citenamefont {Ning}}]{yu2023strain}%
  \BibitemOpen
  \bibfield  {author} {\bibinfo {author} {\bibfnamefont {Y.}~\bibnamefont
  {Yu}}, \bibinfo {author} {\bibfnamefont {C.-D.}\ \bibnamefont {Dong}},
  \bibinfo {author} {\bibfnamefont {R.}~\bibnamefont {Binder}}, \bibinfo
  {author} {\bibfnamefont {S.}~\bibnamefont {Schumacher}},\ and\ \bibinfo
  {author} {\bibfnamefont {C.-Z.}\ \bibnamefont {Ning}},\ }\bibfield  {title}
  {\bibinfo {title} {Strain-induced indirect-to-direct bandgap transition,
  photoluminescence enhancement, and linewidth reduction in bilayer
  \ce{MoTe2}},\ }\href@noop {} {\bibfield  {journal} {\bibinfo  {journal}
  {ACS~Nano}\ }\textbf {\bibinfo {volume} {17}},\ \bibinfo {pages} {4230}
  (\bibinfo {year} {2023})}\BibitemShut {NoStop}%
\bibitem [{\citenamefont {Tebyetekerwa}\ \emph {et~al.}(2020)\citenamefont
  {Tebyetekerwa}, \citenamefont {Zhang}, \citenamefont {Xu}, \citenamefont
  {Truong}, \citenamefont {Yin}, \citenamefont {Lu}, \citenamefont
  {Ramakrishna}, \citenamefont {Macdonald},\ and\ \citenamefont
  {Nguyen}}]{teby+20nano}%
  \BibitemOpen
  \bibfield  {author} {\bibinfo {author} {\bibfnamefont {M.}~\bibnamefont
  {Tebyetekerwa}}, \bibinfo {author} {\bibfnamefont {J.}~\bibnamefont {Zhang}},
  \bibinfo {author} {\bibfnamefont {Z.}~\bibnamefont {Xu}}, \bibinfo {author}
  {\bibfnamefont {T.~N.}\ \bibnamefont {Truong}}, \bibinfo {author}
  {\bibfnamefont {Z.}~\bibnamefont {Yin}}, \bibinfo {author} {\bibfnamefont
  {Y.}~\bibnamefont {Lu}}, \bibinfo {author} {\bibfnamefont {S.}~\bibnamefont
  {Ramakrishna}}, \bibinfo {author} {\bibfnamefont {D.}~\bibnamefont
  {Macdonald}},\ and\ \bibinfo {author} {\bibfnamefont {H.~T.}\ \bibnamefont
  {Nguyen}},\ }\bibfield  {title} {\bibinfo {title} {Mechanisms and
  applications of steady-state photoluminescence spectroscopy in
  two-dimensional transition-metal dichalcogenides},\ }\href@noop {} {\bibfield
   {journal} {\bibinfo  {journal} {ACS~Nano}\ }\textbf {\bibinfo {volume}
  {14}},\ \bibinfo {pages} {14579} (\bibinfo {year} {2020})}\BibitemShut
  {NoStop}%
\bibitem [{\citenamefont {Maniadaki}\ \emph {et~al.}(2016)\citenamefont
  {Maniadaki}, \citenamefont {Kopidakis},\ and\ \citenamefont
  {Remediakis}}]{mani+16ssc}%
  \BibitemOpen
  \bibfield  {author} {\bibinfo {author} {\bibfnamefont {A.~E.}\ \bibnamefont
  {Maniadaki}}, \bibinfo {author} {\bibfnamefont {G.}~\bibnamefont
  {Kopidakis}},\ and\ \bibinfo {author} {\bibfnamefont {I.~N.}\ \bibnamefont
  {Remediakis}},\ }\bibfield  {title} {\bibinfo {title} {Strain engineering of
  electronic properties of transition metal dichalcogenide monolayers},\
  }\href@noop {} {\bibfield  {journal} {\bibinfo  {journal} {Solid State
  Commun.}\ }\textbf {\bibinfo {volume} {227}},\ \bibinfo {pages} {33}
  (\bibinfo {year} {2016})}\BibitemShut {NoStop}%
\bibitem [{\citenamefont {Zhao}\ \emph {et~al.}(2022)\citenamefont {Zhao},
  \citenamefont {Wang}, \citenamefont {Chen},\ and\ \citenamefont
  {Kankala}}]{zhao+22ccr}%
  \BibitemOpen
  \bibfield  {author} {\bibinfo {author} {\bibfnamefont {Y.}~\bibnamefont
  {Zhao}}, \bibinfo {author} {\bibfnamefont {S.-B.}\ \bibnamefont {Wang}},
  \bibinfo {author} {\bibfnamefont {A.-Z.}\ \bibnamefont {Chen}},\ and\
  \bibinfo {author} {\bibfnamefont {R.~K.}\ \bibnamefont {Kankala}},\
  }\bibfield  {title} {\bibinfo {title} {Nanoarchitectured assembly and surface
  of two-dimensional ({2D}) transition metal dichalcogenides (tmdcs) for cancer
  therapy},\ }\href@noop {} {\bibfield  {journal} {\bibinfo  {journal}
  {Coord.~Chem.~Rev.~}\ }\textbf {\bibinfo {volume} {472}},\ \bibinfo {pages}
  {214765} (\bibinfo {year} {2022})}\BibitemShut {NoStop}%
\bibitem [{\citenamefont {Kumar}\ \emph {et~al.}(2024)\citenamefont {Kumar},
  \citenamefont {Mirzaei}, \citenamefont {Kumar}, \citenamefont {Lee},
  \citenamefont {Ghahremani}, \citenamefont {Kim}, \citenamefont {Kim},
  \citenamefont {Kwoka}, \citenamefont {Kumar}, \citenamefont {Kim} \emph
  {et~al.}}]{kumar2024nanoparticles}%
  \BibitemOpen
  \bibfield  {author} {\bibinfo {author} {\bibfnamefont {S.}~\bibnamefont
  {Kumar}}, \bibinfo {author} {\bibfnamefont {A.}~\bibnamefont {Mirzaei}},
  \bibinfo {author} {\bibfnamefont {A.}~\bibnamefont {Kumar}}, \bibinfo
  {author} {\bibfnamefont {M.~H.}\ \bibnamefont {Lee}}, \bibinfo {author}
  {\bibfnamefont {Z.}~\bibnamefont {Ghahremani}}, \bibinfo {author}
  {\bibfnamefont {T.-U.}\ \bibnamefont {Kim}}, \bibinfo {author} {\bibfnamefont
  {J.-Y.}\ \bibnamefont {Kim}}, \bibinfo {author} {\bibfnamefont
  {M.}~\bibnamefont {Kwoka}}, \bibinfo {author} {\bibfnamefont
  {M.}~\bibnamefont {Kumar}}, \bibinfo {author} {\bibfnamefont {S.~S.}\
  \bibnamefont {Kim}}, \emph {et~al.},\ }\bibfield  {title} {\bibinfo {title}
  {Nanoparticles anchored strategy to develop {2D} \ce{MoS2} and \ce{MoSe2}
  based room temperature chemiresistive gas sensors},\ }\href@noop {}
  {\bibfield  {journal} {\bibinfo  {journal} {Coord.~Chem.~Rev.~}\ }\textbf
  {\bibinfo {volume} {503}},\ \bibinfo {pages} {215657} (\bibinfo {year}
  {2024})}\BibitemShut {NoStop}%
\bibitem [{\citenamefont {Wang}\ \emph {et~al.}(2015)\citenamefont {Wang},
  \citenamefont {Wang}, \citenamefont {Wang}, \citenamefont {Wang},
  \citenamefont {Jiang}, \citenamefont {Li}, \citenamefont {Zhang},
  \citenamefont {Zhong},\ and\ \citenamefont {Jiang}}]{wang+15jpcc}%
  \BibitemOpen
  \bibfield  {author} {\bibinfo {author} {\bibfnamefont {H.}~\bibnamefont
  {Wang}}, \bibinfo {author} {\bibfnamefont {X.}~\bibnamefont {Wang}}, \bibinfo
  {author} {\bibfnamefont {L.}~\bibnamefont {Wang}}, \bibinfo {author}
  {\bibfnamefont {J.}~\bibnamefont {Wang}}, \bibinfo {author} {\bibfnamefont
  {D.}~\bibnamefont {Jiang}}, \bibinfo {author} {\bibfnamefont
  {G.}~\bibnamefont {Li}}, \bibinfo {author} {\bibfnamefont {Y.}~\bibnamefont
  {Zhang}}, \bibinfo {author} {\bibfnamefont {H.}~\bibnamefont {Zhong}},\ and\
  \bibinfo {author} {\bibfnamefont {Y.}~\bibnamefont {Jiang}},\ }\bibfield
  {title} {\bibinfo {title} {Phase transition mechanism and electrochemical
  properties of nanocrystalline \ce{MoSe2} as anode materials for the high
  performance lithium-ion battery},\ }\href@noop {} {\bibfield  {journal}
  {\bibinfo  {journal} {J.~Phys.~Chem.~C}\ }\textbf {\bibinfo {volume} {119}},\
  \bibinfo {pages} {10197} (\bibinfo {year} {2015})}\BibitemShut {NoStop}%
\bibitem [{\citenamefont {Bissett}\ \emph {et~al.}(2016)\citenamefont
  {Bissett}, \citenamefont {Worrall}, \citenamefont {Kinloch},\ and\
  \citenamefont {Dryfe}}]{bissett2016comparison}%
  \BibitemOpen
  \bibfield  {author} {\bibinfo {author} {\bibfnamefont {M.~A.}\ \bibnamefont
  {Bissett}}, \bibinfo {author} {\bibfnamefont {S.~D.}\ \bibnamefont
  {Worrall}}, \bibinfo {author} {\bibfnamefont {I.~A.}\ \bibnamefont
  {Kinloch}},\ and\ \bibinfo {author} {\bibfnamefont {R.~A.}\ \bibnamefont
  {Dryfe}},\ }\bibfield  {title} {\bibinfo {title} {Comparison of
  two-dimensional transition metal dichalcogenides for electrochemical
  supercapacitors},\ }\href@noop {} {\bibfield  {journal} {\bibinfo  {journal}
  {Electrochim.~Acta}\ }\textbf {\bibinfo {volume} {201}},\ \bibinfo {pages}
  {30} (\bibinfo {year} {2016})}\BibitemShut {NoStop}%
\bibitem [{\citenamefont {Shen}\ \emph {et~al.}(2019)\citenamefont {Shen},
  \citenamefont {Jiang}, \citenamefont {He}, \citenamefont {Chen},
  \citenamefont {Liu},\ and\ \citenamefont {Zhang}}]{shen+19ns}%
  \BibitemOpen
  \bibfield  {author} {\bibinfo {author} {\bibfnamefont {Q.}~\bibnamefont
  {Shen}}, \bibinfo {author} {\bibfnamefont {P.}~\bibnamefont {Jiang}},
  \bibinfo {author} {\bibfnamefont {H.}~\bibnamefont {He}}, \bibinfo {author}
  {\bibfnamefont {C.}~\bibnamefont {Chen}}, \bibinfo {author} {\bibfnamefont
  {Y.}~\bibnamefont {Liu}},\ and\ \bibinfo {author} {\bibfnamefont
  {M.}~\bibnamefont {Zhang}},\ }\bibfield  {title} {\bibinfo {title}
  {Encapsulation of mose 2 in carbon fibers as anodes for potassium ion
  batteries and nonaqueous battery--supercapacitor hybrid devices},\
  }\href@noop {} {\bibfield  {journal} {\bibinfo  {journal} {Nanoscale}\
  }\textbf {\bibinfo {volume} {11}},\ \bibinfo {pages} {13511} (\bibinfo {year}
  {2019})}\BibitemShut {NoStop}%
\bibitem [{\citenamefont {Chang}\ \emph
  {et~al.}(2014{\natexlab{a}})\citenamefont {Chang}, \citenamefont {Zhang},
  \citenamefont {Zhu}, \citenamefont {Han}, \citenamefont {Pu}, \citenamefont
  {Chang}, \citenamefont {Hsu}, \citenamefont {Huang}, \citenamefont {Hsu},
  \citenamefont {Chiu} \emph {et~al.}}]{chan+14nano}%
  \BibitemOpen
  \bibfield  {author} {\bibinfo {author} {\bibfnamefont {Y.-H.}\ \bibnamefont
  {Chang}}, \bibinfo {author} {\bibfnamefont {W.}~\bibnamefont {Zhang}},
  \bibinfo {author} {\bibfnamefont {Y.}~\bibnamefont {Zhu}}, \bibinfo {author}
  {\bibfnamefont {Y.}~\bibnamefont {Han}}, \bibinfo {author} {\bibfnamefont
  {J.}~\bibnamefont {Pu}}, \bibinfo {author} {\bibfnamefont {J.-K.}\
  \bibnamefont {Chang}}, \bibinfo {author} {\bibfnamefont {W.-T.}\ \bibnamefont
  {Hsu}}, \bibinfo {author} {\bibfnamefont {J.-K.}\ \bibnamefont {Huang}},
  \bibinfo {author} {\bibfnamefont {C.-L.}\ \bibnamefont {Hsu}}, \bibinfo
  {author} {\bibfnamefont {M.-H.}\ \bibnamefont {Chiu}}, \emph {et~al.},\
  }\bibfield  {title} {\bibinfo {title} {Monolayer \ce{MoSe2} grown by chemical
  vapor deposition for fast photodetection},\ }\href@noop {} {\bibfield
  {journal} {\bibinfo  {journal} {ACS~Nano}\ }\textbf {\bibinfo {volume} {8}},\
  \bibinfo {pages} {8582} (\bibinfo {year} {2014}{\natexlab{a}})}\BibitemShut
  {NoStop}%
\bibitem [{\citenamefont {Blauth}\ \emph {et~al.}(2019)\citenamefont {Blauth},
  \citenamefont {Vest}, \citenamefont {Rosemary}, \citenamefont {Prechtl},
  \citenamefont {Hartwig}, \citenamefont {J\"{u}rgensen}, \citenamefont
  {Kaniber}, \citenamefont {Stier},\ and\ \citenamefont
  {Finley}}]{blau+19acsphotonics}%
  \BibitemOpen
  \bibfield  {author} {\bibinfo {author} {\bibfnamefont {M.}~\bibnamefont
  {Blauth}}, \bibinfo {author} {\bibfnamefont {G.}~\bibnamefont {Vest}},
  \bibinfo {author} {\bibfnamefont {S.~L.}\ \bibnamefont {Rosemary}}, \bibinfo
  {author} {\bibfnamefont {M.}~\bibnamefont {Prechtl}}, \bibinfo {author}
  {\bibfnamefont {O.}~\bibnamefont {Hartwig}}, \bibinfo {author} {\bibfnamefont
  {M.}~\bibnamefont {J\"{u}rgensen}}, \bibinfo {author} {\bibfnamefont
  {M.}~\bibnamefont {Kaniber}}, \bibinfo {author} {\bibfnamefont {A.~V.}\
  \bibnamefont {Stier}},\ and\ \bibinfo {author} {\bibfnamefont {J.~J.}\
  \bibnamefont {Finley}},\ }\bibfield  {title} {\bibinfo {title} {Ultracompact
  photodetection in atomically thin \ce{MoSe2}},\ }\href@noop {} {\bibfield
  {journal} {\bibinfo  {journal} {ACS~Photon.}\ }\textbf {\bibinfo {volume}
  {6}},\ \bibinfo {pages} {1902} (\bibinfo {year} {2019})}\BibitemShut
  {NoStop}%
\bibitem [{\citenamefont {Li}\ \emph {et~al.}(2018)\citenamefont {Li},
  \citenamefont {Huang}, \citenamefont {Zhang}, \citenamefont {Zhu},
  \citenamefont {Li}, \citenamefont {Meng},\ and\ \citenamefont
  {Tian}}]{li+18cpl}%
  \BibitemOpen
  \bibfield  {author} {\bibinfo {author} {\bibfnamefont {H.}~\bibnamefont
  {Li}}, \bibinfo {author} {\bibfnamefont {S.}~\bibnamefont {Huang}}, \bibinfo
  {author} {\bibfnamefont {Q.}~\bibnamefont {Zhang}}, \bibinfo {author}
  {\bibfnamefont {Z.}~\bibnamefont {Zhu}}, \bibinfo {author} {\bibfnamefont
  {C.}~\bibnamefont {Li}}, \bibinfo {author} {\bibfnamefont {J.}~\bibnamefont
  {Meng}},\ and\ \bibinfo {author} {\bibfnamefont {Y.}~\bibnamefont {Tian}},\
  }\bibfield  {title} {\bibinfo {title} {Nonmetal doping induced electronic and
  magnetic properties in \ce{MoSe2} monolayer},\ }\href@noop {} {\bibfield
  {journal} {\bibinfo  {journal} {Chem.~Phys.~Lett.~}\ }\textbf {\bibinfo
  {volume} {692}},\ \bibinfo {pages} {69} (\bibinfo {year} {2018})}\BibitemShut
  {NoStop}%
\bibitem [{\citenamefont {Zhang}\ \emph {et~al.}(2019)\citenamefont {Zhang},
  \citenamefont {Wong}, \citenamefont {Zhou}, \citenamefont {Rath},
  \citenamefont {Huang}, \citenamefont {Wang}, \citenamefont {Morton},
  \citenamefont {Yuan}, \citenamefont {Zhang}, \citenamefont {Chua} \emph
  {et~al.}}]{zhan+19nano}%
  \BibitemOpen
  \bibfield  {author} {\bibinfo {author} {\bibfnamefont {W.}~\bibnamefont
  {Zhang}}, \bibinfo {author} {\bibfnamefont {P.~K.~J.}\ \bibnamefont {Wong}},
  \bibinfo {author} {\bibfnamefont {X.}~\bibnamefont {Zhou}}, \bibinfo {author}
  {\bibfnamefont {A.}~\bibnamefont {Rath}}, \bibinfo {author} {\bibfnamefont
  {Z.}~\bibnamefont {Huang}}, \bibinfo {author} {\bibfnamefont
  {H.}~\bibnamefont {Wang}}, \bibinfo {author} {\bibfnamefont {S.~A.}\
  \bibnamefont {Morton}}, \bibinfo {author} {\bibfnamefont {J.}~\bibnamefont
  {Yuan}}, \bibinfo {author} {\bibfnamefont {L.}~\bibnamefont {Zhang}},
  \bibinfo {author} {\bibfnamefont {R.}~\bibnamefont {Chua}}, \emph {et~al.},\
  }\bibfield  {title} {\bibinfo {title} {Ferromagnet/two-dimensional
  semiconducting transition-metal dichalcogenide interface with perpendicular
  magnetic anisotropy},\ }\href@noop {} {\bibfield  {journal} {\bibinfo
  {journal} {ACS~Nano}\ }\textbf {\bibinfo {volume} {13}},\ \bibinfo {pages}
  {2253} (\bibinfo {year} {2019})}\BibitemShut {NoStop}%
\bibitem [{\citenamefont {Khan}\ \emph {et~al.}(2020)\citenamefont {Khan},
  \citenamefont {Rehman}, \citenamefont {Rehman}, \citenamefont {Rehman~Sagar},
  \citenamefont {Kim}, \citenamefont {Waseem~Khalil}, \citenamefont {Shinde},
  \citenamefont {Sharma}, \citenamefont {Eom}, \citenamefont {Chan~Jun} \emph
  {et~al.}}]{khan+20aplmater}%
  \BibitemOpen
  \bibfield  {author} {\bibinfo {author} {\bibfnamefont {M.~F.}\ \bibnamefont
  {Khan}}, \bibinfo {author} {\bibfnamefont {S.}~\bibnamefont {Rehman}},
  \bibinfo {author} {\bibfnamefont {M.~A.}\ \bibnamefont {Rehman}}, \bibinfo
  {author} {\bibfnamefont {R.~U.}\ \bibnamefont {Rehman~Sagar}}, \bibinfo
  {author} {\bibfnamefont {D.-k.}\ \bibnamefont {Kim}}, \bibinfo {author}
  {\bibfnamefont {H.}~\bibnamefont {Waseem~Khalil}}, \bibinfo {author}
  {\bibfnamefont {P.~A.}\ \bibnamefont {Shinde}}, \bibinfo {author}
  {\bibfnamefont {P.~R.}\ \bibnamefont {Sharma}}, \bibinfo {author}
  {\bibfnamefont {J.}~\bibnamefont {Eom}}, \bibinfo {author} {\bibfnamefont
  {S.}~\bibnamefont {Chan~Jun}}, \emph {et~al.},\ }\bibfield  {title} {\bibinfo
  {title} {Multi-heterostructured spin-valve junction of vertical
  flg/mose$_2$flg},\ }\href@noop {} {\bibfield  {journal} {\bibinfo  {journal}
  {APL Mater.}\ }\textbf {\bibinfo {volume} {8}},\ \bibinfo {pages} {071104}
  (\bibinfo {year} {2020})}\BibitemShut {NoStop}%
\bibitem [{\citenamefont {Huang}\ \emph {et~al.}(2024)\citenamefont {Huang},
  \citenamefont {Gu}, \citenamefont {Fang},\ and\ \citenamefont
  {Liu}}]{huang2024substrate}%
  \BibitemOpen
  \bibfield  {author} {\bibinfo {author} {\bibfnamefont {L.}~\bibnamefont
  {Huang}}, \bibinfo {author} {\bibfnamefont {H.}~\bibnamefont {Gu}}, \bibinfo
  {author} {\bibfnamefont {M.}~\bibnamefont {Fang}},\ and\ \bibinfo {author}
  {\bibfnamefont {S.}~\bibnamefont {Liu}},\ }\bibfield  {title} {\bibinfo
  {title} {Substrate-tuned dielectric screening effect on optical properties of
  monolayer \ce{MoSe2}},\ }\href@noop {} {\bibfield  {journal} {\bibinfo
  {journal} {Appl.~Surf.~Sci.~}\ }\textbf {\bibinfo {volume} {644}},\ \bibinfo
  {pages} {158748} (\bibinfo {year} {2024})}\BibitemShut {NoStop}%
\bibitem [{\citenamefont {Valouch}\ \emph {et~al.}(2022)\citenamefont
  {Valouch}, \citenamefont {Hermes}, \citenamefont {Hengen}, \citenamefont
  {Send},\ and\ \citenamefont {Bruder}}]{patent_detector}%
  \BibitemOpen
  \bibfield  {author} {\bibinfo {author} {\bibfnamefont {S.}~\bibnamefont
  {Valouch}}, \bibinfo {author} {\bibfnamefont {W.}~\bibnamefont {Hermes}},
  \bibinfo {author} {\bibfnamefont {S.}~\bibnamefont {Hengen}}, \bibinfo
  {author} {\bibfnamefont {R.}~\bibnamefont {Send}},\ and\ \bibinfo {author}
  {\bibfnamefont {I.}~\bibnamefont {Bruder}},\ }\href@noop {} {\bibinfo {title}
  {Detector for an optical detection of at least one object}} (\bibinfo {year}
  {2022}),\ \bibinfo {note} {{United States Patent No.
  US11428787B2}}\BibitemShut {NoStop}%
\bibitem [{\citenamefont {Lin}\ \emph {et~al.}(2022)\citenamefont {Lin},
  \citenamefont {Manipatruni}, \citenamefont {Gosavi}, \citenamefont {Chang},
  \citenamefont {Nikonov},\ and\ \citenamefont {A.~Young}}]{patent_transistor}%
  \BibitemOpen
  \bibfield  {author} {\bibinfo {author} {\bibfnamefont {C.-C.}\ \bibnamefont
  {Lin}}, \bibinfo {author} {\bibfnamefont {S.}~\bibnamefont {Manipatruni}},
  \bibinfo {author} {\bibfnamefont {T.}~\bibnamefont {Gosavi}}, \bibinfo
  {author} {\bibfnamefont {S.-C.}\ \bibnamefont {Chang}}, \bibinfo {author}
  {\bibfnamefont {D.}~\bibnamefont {Nikonov}},\ and\ \bibinfo {author}
  {\bibfnamefont {I.}~\bibnamefont {A.~Young}},\ }\href@noop {} {\bibinfo
  {title} {Magnetoelectric spin orbit logic transistor with a spin filter}}
  (\bibinfo {year} {2022}),\ \bibinfo {note} {{United States Patent No.
  US11398562B2}}\BibitemShut {NoStop}%
\bibitem [{\citenamefont {Choi}\ and\ \citenamefont
  {CHA}(2022)}]{patent_battery}%
  \BibitemOpen
  \bibfield  {author} {\bibinfo {author} {\bibfnamefont {W.}~\bibnamefont
  {Choi}}\ and\ \bibinfo {author} {\bibfnamefont {E.}~\bibnamefont {CHA}},\
  }\href@noop {} {\bibinfo {title} {Passivation of lithium metal by
  two-dimensional materials for rechargeable batteries}} (\bibinfo {year}
  {2022}),\ \bibinfo {note} {{United States Patent No.
  US11355739B2}}\BibitemShut {NoStop}%
\bibitem [{\citenamefont {V}\ and\ \citenamefont
  {E}(2024)}]{patent_conversion}%
  \BibitemOpen
  \bibfield  {author} {\bibinfo {author} {\bibfnamefont {K.~G.}\ \bibnamefont
  {V}}\ and\ \bibinfo {author} {\bibfnamefont {R.~S.}\ \bibnamefont {E}},\
  }\href@noop {} {\bibinfo {title} {Exciton polariton optical interconnect}}
  (\bibinfo {year} {2024}),\ \bibinfo {note} {{United States Patent No.
  US11874581B2}}\BibitemShut {NoStop}%
\bibitem [{\citenamefont {Thibado}(2023)}]{patent_energy}%
  \BibitemOpen
  \bibfield  {author} {\bibinfo {author} {\bibfnamefont {P.}~\bibnamefont
  {Thibado}},\ }\href@noop {} {\bibinfo {title} {Energy harvesting devices and
  sensors, and methods of making and use thereof}} (\bibinfo {year} {2023}),\
  \bibinfo {note} {{Japan Patent No. JP7294682B2}}\BibitemShut {NoStop}%
\bibitem [{\citenamefont {Deng}\ \emph {et~al.}(2018)\citenamefont {Deng},
  \citenamefont {Li},\ and\ \citenamefont {Li}}]{deng+18pe}%
  \BibitemOpen
  \bibfield  {author} {\bibinfo {author} {\bibfnamefont {S.}~\bibnamefont
  {Deng}}, \bibinfo {author} {\bibfnamefont {L.}~\bibnamefont {Li}},\ and\
  \bibinfo {author} {\bibfnamefont {M.}~\bibnamefont {Li}},\ }\bibfield
  {title} {\bibinfo {title} {Stability of direct band gap under mechanical
  strains for monolayer \ce{MoS2}, \ce{MoSe2}, \ce{WS2} and \ce{WSe2}},\
  }\href@noop {} {\bibfield  {journal} {\bibinfo  {journal} {Physica E}\
  }\textbf {\bibinfo {volume} {101}},\ \bibinfo {pages} {44} (\bibinfo {year}
  {2018})}\BibitemShut {NoStop}%
\bibitem [{\citenamefont {Blundo}\ \emph {et~al.}(2022)\citenamefont {Blundo},
  \citenamefont {Surrente}, \citenamefont {Spirito}, \citenamefont {Pettinari},
  \citenamefont {Yildirim}, \citenamefont {Chavarin}, \citenamefont
  {Baldassarre}, \citenamefont {Felici},\ and\ \citenamefont
  {Polimeni}}]{blun+22nl}%
  \BibitemOpen
  \bibfield  {author} {\bibinfo {author} {\bibfnamefont {E.}~\bibnamefont
  {Blundo}}, \bibinfo {author} {\bibfnamefont {A.}~\bibnamefont {Surrente}},
  \bibinfo {author} {\bibfnamefont {D.}~\bibnamefont {Spirito}}, \bibinfo
  {author} {\bibfnamefont {G.}~\bibnamefont {Pettinari}}, \bibinfo {author}
  {\bibfnamefont {T.}~\bibnamefont {Yildirim}}, \bibinfo {author}
  {\bibfnamefont {C.~A.}\ \bibnamefont {Chavarin}}, \bibinfo {author}
  {\bibfnamefont {L.}~\bibnamefont {Baldassarre}}, \bibinfo {author}
  {\bibfnamefont {M.}~\bibnamefont {Felici}},\ and\ \bibinfo {author}
  {\bibfnamefont {A.}~\bibnamefont {Polimeni}},\ }\bibfield  {title} {\bibinfo
  {title} {Vibrational properties in highly strained hexagonal boron nitride
  bubbles},\ }\href@noop {} {\bibfield  {journal} {\bibinfo  {journal}
  {Nano~Lett.~}\ }\textbf {\bibinfo {volume} {22}},\ \bibinfo {pages} {1525}
  (\bibinfo {year} {2022})}\BibitemShut {NoStop}%
\bibitem [{\citenamefont {Ochoa}\ \emph {et~al.}(2017)\citenamefont {Ochoa},
  \citenamefont {Zarzuela},\ and\ \citenamefont {Tserkovnyak}}]{ocho+17prl}%
  \BibitemOpen
  \bibfield  {author} {\bibinfo {author} {\bibfnamefont {H.}~\bibnamefont
  {Ochoa}}, \bibinfo {author} {\bibfnamefont {R.}~\bibnamefont {Zarzuela}},\
  and\ \bibinfo {author} {\bibfnamefont {Y.}~\bibnamefont {Tserkovnyak}},\
  }\bibfield  {title} {\bibinfo {title} {Emergent gauge fields from curvature
  in single layers of transition-metal dichalcogenides},\ }\href@noop {}
  {\bibfield  {journal} {\bibinfo  {journal} {Phys.~Rev.~Lett.}\ }\textbf
  {\bibinfo {volume} {118}},\ \bibinfo {pages} {026801} (\bibinfo {year}
  {2017})}\BibitemShut {NoStop}%
\bibitem [{\citenamefont {Castellanos-Gomez}\ \emph {et~al.}(2013)\citenamefont
  {Castellanos-Gomez}, \citenamefont {Rold{\'a}n}, \citenamefont {Cappelluti},
  \citenamefont {Buscema}, \citenamefont {Guinea}, \citenamefont {Van
  Der~Zant},\ and\ \citenamefont {Steele}}]{castellanos2013local}%
  \BibitemOpen
  \bibfield  {author} {\bibinfo {author} {\bibfnamefont {A.}~\bibnamefont
  {Castellanos-Gomez}}, \bibinfo {author} {\bibfnamefont {R.}~\bibnamefont
  {Rold{\'a}n}}, \bibinfo {author} {\bibfnamefont {E.}~\bibnamefont
  {Cappelluti}}, \bibinfo {author} {\bibfnamefont {M.}~\bibnamefont {Buscema}},
  \bibinfo {author} {\bibfnamefont {F.}~\bibnamefont {Guinea}}, \bibinfo
  {author} {\bibfnamefont {H.~S.}\ \bibnamefont {Van Der~Zant}},\ and\ \bibinfo
  {author} {\bibfnamefont {G.~A.}\ \bibnamefont {Steele}},\ }\bibfield  {title}
  {\bibinfo {title} {Local strain engineering in atomically thin \ce{MoS2}},\
  }\href@noop {} {\bibfield  {journal} {\bibinfo  {journal} {Nano letters}\
  }\textbf {\bibinfo {volume} {13}},\ \bibinfo {pages} {5361} (\bibinfo {year}
  {2013})}\BibitemShut {NoStop}%
\bibitem [{\citenamefont {Liu}\ \emph {et~al.}(2020)\citenamefont {Liu},
  \citenamefont {Sachan}, \citenamefont {Howell}, \citenamefont {Conde-Rubio},
  \citenamefont {Knoll}, \citenamefont {Boero}, \citenamefont {Zenobi},\ and\
  \citenamefont {Brugger}}]{liu+20nl}%
  \BibitemOpen
  \bibfield  {author} {\bibinfo {author} {\bibfnamefont {X.}~\bibnamefont
  {Liu}}, \bibinfo {author} {\bibfnamefont {A.~K.}\ \bibnamefont {Sachan}},
  \bibinfo {author} {\bibfnamefont {S.~T.}\ \bibnamefont {Howell}}, \bibinfo
  {author} {\bibfnamefont {A.}~\bibnamefont {Conde-Rubio}}, \bibinfo {author}
  {\bibfnamefont {A.~W.}\ \bibnamefont {Knoll}}, \bibinfo {author}
  {\bibfnamefont {G.}~\bibnamefont {Boero}}, \bibinfo {author} {\bibfnamefont
  {R.}~\bibnamefont {Zenobi}},\ and\ \bibinfo {author} {\bibfnamefont
  {J.}~\bibnamefont {Brugger}},\ }\bibfield  {title} {\bibinfo {title}
  {Thermomechanical nanostraining of two-dimensional materials},\ }\href@noop
  {} {\bibfield  {journal} {\bibinfo  {journal} {Nano~Lett.~}\ }\textbf
  {\bibinfo {volume} {20}},\ \bibinfo {pages} {8250} (\bibinfo {year}
  {2020})}\BibitemShut {NoStop}%
\bibitem [{\citenamefont {Wang}\ \emph {et~al.}(2021)\citenamefont {Wang},
  \citenamefont {Maisch}, \citenamefont {Tang}, \citenamefont {Zhao},
  \citenamefont {Yang}, \citenamefont {Joos}, \citenamefont {Portalupi},
  \citenamefont {Michler},\ and\ \citenamefont {Smet}}]{wang+21nl}%
  \BibitemOpen
  \bibfield  {author} {\bibinfo {author} {\bibfnamefont {Q.}~\bibnamefont
  {Wang}}, \bibinfo {author} {\bibfnamefont {J.}~\bibnamefont {Maisch}},
  \bibinfo {author} {\bibfnamefont {F.}~\bibnamefont {Tang}}, \bibinfo {author}
  {\bibfnamefont {D.}~\bibnamefont {Zhao}}, \bibinfo {author} {\bibfnamefont
  {S.}~\bibnamefont {Yang}}, \bibinfo {author} {\bibfnamefont {R.}~\bibnamefont
  {Joos}}, \bibinfo {author} {\bibfnamefont {S.~L.}\ \bibnamefont {Portalupi}},
  \bibinfo {author} {\bibfnamefont {P.}~\bibnamefont {Michler}},\ and\ \bibinfo
  {author} {\bibfnamefont {J.~H.}\ \bibnamefont {Smet}},\ }\bibfield  {title}
  {\bibinfo {title} {Highly polarized single photons from strain-induced
  quasi-{1D} localized excitons in \ce{WSe2}},\ }\href
  {https://doi.org/10.1021/acs.nanolett.1c01927} {\bibfield  {journal}
  {\bibinfo  {journal} {Nano~Lett.~}\ }\textbf {\bibinfo {volume} {21}},\
  \bibinfo {pages} {7175} (\bibinfo {year} {2021})}\BibitemShut {NoStop}%
\bibitem [{\citenamefont {Liu}\ \emph {et~al.}(2022)\citenamefont {Liu},
  \citenamefont {Lu}, \citenamefont {Guan},\ and\ \citenamefont
  {Hu}}]{liu+22acsami}%
  \BibitemOpen
  \bibfield  {author} {\bibinfo {author} {\bibfnamefont {J.}~\bibnamefont
  {Liu}}, \bibinfo {author} {\bibfnamefont {N.}~\bibnamefont {Lu}}, \bibinfo
  {author} {\bibfnamefont {J.}~\bibnamefont {Guan}},\ and\ \bibinfo {author}
  {\bibfnamefont {Y.}~\bibnamefont {Hu}},\ }\bibfield  {title} {\bibinfo
  {title} {Laser shock-induced nano-twist of transition metal
  dichalcogenides},\ }\href@noop {} {\bibfield  {journal} {\bibinfo  {journal}
  {ACS~Appl.~Mater.~Interfaces}\ }\textbf {\bibinfo {volume} {14}},\ \bibinfo
  {pages} {37213} (\bibinfo {year} {2022})}\BibitemShut {NoStop}%
\bibitem [{\citenamefont {Hohenberg}\ and\ \citenamefont
  {Kohn}(1964)}]{hohenberg1964inhomogeneous}%
  \BibitemOpen
  \bibfield  {author} {\bibinfo {author} {\bibfnamefont {P.}~\bibnamefont
  {Hohenberg}}\ and\ \bibinfo {author} {\bibfnamefont {W.}~\bibnamefont
  {Kohn}},\ }\bibfield  {title} {\bibinfo {title} {Inhomogeneous electron
  gas},\ }\href@noop {} {\bibfield  {journal} {\bibinfo  {journal}
  {Phys.~Rev.}\ }\textbf {\bibinfo {volume} {136}},\ \bibinfo {pages} {B864}
  (\bibinfo {year} {1964})}\BibitemShut {NoStop}%
\bibitem [{\citenamefont {Kohn}\ and\ \citenamefont
  {Sham}(1965)}]{kohn1965self}%
  \BibitemOpen
  \bibfield  {author} {\bibinfo {author} {\bibfnamefont {W.}~\bibnamefont
  {Kohn}}\ and\ \bibinfo {author} {\bibfnamefont {L.~J.}\ \bibnamefont
  {Sham}},\ }\bibfield  {title} {\bibinfo {title} {Self-consistent equations
  including exchange and correlation effects},\ }\href@noop {} {\bibfield
  {journal} {\bibinfo  {journal} {Phys.~Rev.}\ }\textbf {\bibinfo {volume}
  {140}},\ \bibinfo {pages} {A1133} (\bibinfo {year} {1965})}\BibitemShut
  {NoStop}%
\bibitem [{\citenamefont {Burke}(2012)}]{burk12jcp}%
  \BibitemOpen
  \bibfield  {author} {\bibinfo {author} {\bibfnamefont {K.}~\bibnamefont
  {Burke}},\ }\bibfield  {title} {\bibinfo {title} {Perspective on density
  functional theory},\ }\href@noop {} {\bibfield  {journal} {\bibinfo
  {journal} {J.~Chem.~Phys.~}\ }\textbf {\bibinfo {volume} {136}} (\bibinfo
  {year} {2012})}\BibitemShut {NoStop}%
\bibitem [{\citenamefont {Popescu}\ and\ \citenamefont
  {Zunger}(2012)}]{popescu_zunger}%
  \BibitemOpen
  \bibfield  {author} {\bibinfo {author} {\bibfnamefont {V.}~\bibnamefont
  {Popescu}}\ and\ \bibinfo {author} {\bibfnamefont {A.}~\bibnamefont
  {Zunger}},\ }\bibfield  {title} {\bibinfo {title} {Extracting \textit{E}
  versus $\vec{k}$ effective band structure from supercell calculations on
  alloys and impurities},\ }\href {https://doi.org/10.1103/PhysRevB.85.085201}
  {\bibfield  {journal} {\bibinfo  {journal} {Phys. Rev. B}\ }\textbf {\bibinfo
  {volume} {85}},\ \bibinfo {pages} {085201} (\bibinfo {year}
  {2012})}\BibitemShut {NoStop}%
\bibitem [{\citenamefont {Boykin}\ and\ \citenamefont
  {Klimeck}(2005)}]{boyk-klim05prb}%
  \BibitemOpen
  \bibfield  {author} {\bibinfo {author} {\bibfnamefont {T.~B.}\ \bibnamefont
  {Boykin}}\ and\ \bibinfo {author} {\bibfnamefont {G.}~\bibnamefont
  {Klimeck}},\ }\bibfield  {title} {\bibinfo {title} {Practical application of
  zone-folding concepts in tight-binding calculations},\ }\href@noop {}
  {\bibfield  {journal} {\bibinfo  {journal} {Phys.~Rev.~B}\ }\textbf {\bibinfo
  {volume} {71}},\ \bibinfo {pages} {115215} (\bibinfo {year}
  {2005})}\BibitemShut {NoStop}%
\bibitem [{\citenamefont {Boykin}\ \emph {et~al.}(2007)\citenamefont {Boykin},
  \citenamefont {Kharche}, \citenamefont {Klimeck},\ and\ \citenamefont
  {Korkusinski}}]{boyk+07jpcm}%
  \BibitemOpen
  \bibfield  {author} {\bibinfo {author} {\bibfnamefont {T.~B.}\ \bibnamefont
  {Boykin}}, \bibinfo {author} {\bibfnamefont {N.}~\bibnamefont {Kharche}},
  \bibinfo {author} {\bibfnamefont {G.}~\bibnamefont {Klimeck}},\ and\ \bibinfo
  {author} {\bibfnamefont {M.}~\bibnamefont {Korkusinski}},\ }\bibfield
  {title} {\bibinfo {title} {Approximate bandstructures of semiconductor alloys
  from tight-binding supercell calculations},\ }\href@noop {} {\bibfield
  {journal} {\bibinfo  {journal} {J.~Phys.:~Condens.~Matter.~}\ }\textbf
  {\bibinfo {volume} {19}},\ \bibinfo {pages} {036203} (\bibinfo {year}
  {2007})}\BibitemShut {NoStop}%
\bibitem [{\citenamefont {Krumland}\ and\ \citenamefont
  {Cocchi}(2021)}]{krumland2021conditions}%
  \BibitemOpen
  \bibfield  {author} {\bibinfo {author} {\bibfnamefont {J.}~\bibnamefont
  {Krumland}}\ and\ \bibinfo {author} {\bibfnamefont {C.}~\bibnamefont
  {Cocchi}},\ }\bibfield  {title} {\bibinfo {title} {Conditions for electronic
  hybridization between transition-metal dichalcogenide monolayers and
  physisorbed carbon-conjugated molecules},\ }\href@noop {} {\bibfield
  {journal} {\bibinfo  {journal} {Electron.~Struct.}\ }\textbf {\bibinfo
  {volume} {3}},\ \bibinfo {pages} {044003} (\bibinfo {year}
  {2021})}\BibitemShut {NoStop}%
\bibitem [{\citenamefont {Krumland}\ and\ \citenamefont
  {Cocchi}(2024)}]{krum-cocc23pssa}%
  \BibitemOpen
  \bibfield  {author} {\bibinfo {author} {\bibfnamefont {J.}~\bibnamefont
  {Krumland}}\ and\ \bibinfo {author} {\bibfnamefont {C.}~\bibnamefont
  {Cocchi}},\ }\bibfield  {title} {\bibinfo {title} {Electronic structure of
  low-dimensional inorganic/organic interfaces: Hybrid density functional
  theory, g 0 w 0, and electrostatic models},\ }\href@noop {} {\bibfield
  {journal} {\bibinfo  {journal} {Phys.~Status~Solidi~A}\ }\textbf {\bibinfo
  {volume} {221}},\ \bibinfo {pages} {2300089} (\bibinfo {year}
  {2024})}\BibitemShut {NoStop}%
\bibitem [{\citenamefont {Giannozzi}\ \emph {et~al.}(2009)\citenamefont
  {Giannozzi}, \citenamefont {Baroni}, \citenamefont {Bonini}, \citenamefont
  {Calandra}, \citenamefont {Car}, \citenamefont {Cavazzoni}, \citenamefont
  {Ceresoli}, \citenamefont {Chiarotti}, \citenamefont {Cococcioni},
  \citenamefont {Dabo} \emph {et~al.}}]{giannozzi2009quantum}%
  \BibitemOpen
  \bibfield  {author} {\bibinfo {author} {\bibfnamefont {P.}~\bibnamefont
  {Giannozzi}}, \bibinfo {author} {\bibfnamefont {S.}~\bibnamefont {Baroni}},
  \bibinfo {author} {\bibfnamefont {N.}~\bibnamefont {Bonini}}, \bibinfo
  {author} {\bibfnamefont {M.}~\bibnamefont {Calandra}}, \bibinfo {author}
  {\bibfnamefont {R.}~\bibnamefont {Car}}, \bibinfo {author} {\bibfnamefont
  {C.}~\bibnamefont {Cavazzoni}}, \bibinfo {author} {\bibfnamefont
  {D.}~\bibnamefont {Ceresoli}}, \bibinfo {author} {\bibfnamefont {G.~L.}\
  \bibnamefont {Chiarotti}}, \bibinfo {author} {\bibfnamefont {M.}~\bibnamefont
  {Cococcioni}}, \bibinfo {author} {\bibfnamefont {I.}~\bibnamefont {Dabo}},
  \emph {et~al.},\ }\bibfield  {title} {\bibinfo {title} {{QUANTUM ESPRESSO}: a
  modular and open-source software project for quantum simulations of
  materials},\ }\href@noop {} {\bibfield  {journal} {\bibinfo  {journal}
  {J.~Phys.:~Condens.~Matter.~}\ }\textbf {\bibinfo {volume} {21}},\ \bibinfo
  {pages} {395502} (\bibinfo {year} {2009})}\BibitemShut {NoStop}%
\bibitem [{\citenamefont {Giannozzi}\ \emph {et~al.}(2017)\citenamefont
  {Giannozzi}, \citenamefont {Andreussi}, \citenamefont {Brumme}, \citenamefont
  {Bunau}, \citenamefont {Nardelli}, \citenamefont {Calandra}, \citenamefont
  {Car}, \citenamefont {Cavazzoni}, \citenamefont {Ceresoli}, \citenamefont
  {Cococcioni} \emph {et~al.}}]{giannozzi2017advanced}%
  \BibitemOpen
  \bibfield  {author} {\bibinfo {author} {\bibfnamefont {P.}~\bibnamefont
  {Giannozzi}}, \bibinfo {author} {\bibfnamefont {O.}~\bibnamefont
  {Andreussi}}, \bibinfo {author} {\bibfnamefont {T.}~\bibnamefont {Brumme}},
  \bibinfo {author} {\bibfnamefont {O.}~\bibnamefont {Bunau}}, \bibinfo
  {author} {\bibfnamefont {M.~B.}\ \bibnamefont {Nardelli}}, \bibinfo {author}
  {\bibfnamefont {M.}~\bibnamefont {Calandra}}, \bibinfo {author}
  {\bibfnamefont {R.}~\bibnamefont {Car}}, \bibinfo {author} {\bibfnamefont
  {C.}~\bibnamefont {Cavazzoni}}, \bibinfo {author} {\bibfnamefont
  {D.}~\bibnamefont {Ceresoli}}, \bibinfo {author} {\bibfnamefont
  {M.}~\bibnamefont {Cococcioni}}, \emph {et~al.},\ }\bibfield  {title}
  {\bibinfo {title} {Advanced capabilities for materials modelling with
  {Quantum ESPRESSO}},\ }\href@noop {} {\bibfield  {journal} {\bibinfo
  {journal} {J.~Phys.:~Condens.~Matter.~}\ }\textbf {\bibinfo {volume} {29}},\
  \bibinfo {pages} {465901} (\bibinfo {year} {2017})}\BibitemShut {NoStop}%
\bibitem [{\citenamefont {Perdew}\ \emph {et~al.}(1996)\citenamefont {Perdew},
  \citenamefont {Burke},\ and\ \citenamefont {Wang}}]{perdew1996generalized}%
  \BibitemOpen
  \bibfield  {author} {\bibinfo {author} {\bibfnamefont {J.~P.}\ \bibnamefont
  {Perdew}}, \bibinfo {author} {\bibfnamefont {K.}~\bibnamefont {Burke}},\ and\
  \bibinfo {author} {\bibfnamefont {Y.}~\bibnamefont {Wang}},\ }\bibfield
  {title} {\bibinfo {title} {Generalized gradient approximation for the
  exchange-correlation hole of a many-electron system},\ }\href@noop {}
  {\bibfield  {journal} {\bibinfo  {journal} {Phys.~Rev.~B}\ }\textbf {\bibinfo
  {volume} {54}},\ \bibinfo {pages} {16533} (\bibinfo {year}
  {1996})}\BibitemShut {NoStop}%
\bibitem [{\citenamefont {Schlipf}\ and\ \citenamefont
  {Gygi}(2015)}]{schlipf2015optimization}%
  \BibitemOpen
  \bibfield  {author} {\bibinfo {author} {\bibfnamefont {M.}~\bibnamefont
  {Schlipf}}\ and\ \bibinfo {author} {\bibfnamefont {F.}~\bibnamefont {Gygi}},\
  }\bibfield  {title} {\bibinfo {title} {Optimization algorithm for the
  generation of {ONCV} pseudopotentials},\ }\href@noop {} {\bibfield  {journal}
  {\bibinfo  {journal} {Comput.~Phys.~Commun.~}\ }\textbf {\bibinfo {volume}
  {196}},\ \bibinfo {pages} {36} (\bibinfo {year} {2015})}\BibitemShut
  {NoStop}%
\bibitem [{\citenamefont {Nepal}\ \emph {et~al.}(2019)\citenamefont {Nepal},
  \citenamefont {Yu}, \citenamefont {Yan},\ and\ \citenamefont
  {Ruzsinszky}}]{nepal2019first}%
  \BibitemOpen
  \bibfield  {author} {\bibinfo {author} {\bibfnamefont {N.~K.}\ \bibnamefont
  {Nepal}}, \bibinfo {author} {\bibfnamefont {L.}~\bibnamefont {Yu}}, \bibinfo
  {author} {\bibfnamefont {Q.}~\bibnamefont {Yan}},\ and\ \bibinfo {author}
  {\bibfnamefont {A.}~\bibnamefont {Ruzsinszky}},\ }\bibfield  {title}
  {\bibinfo {title} {First-principles study of mechanical and electronic
  properties of bent monolayer transition metal dichalcogenides},\ }\href@noop
  {} {\bibfield  {journal} {\bibinfo  {journal} {Phys.~Rev.~Mater.}\ }\textbf
  {\bibinfo {volume} {3}},\ \bibinfo {pages} {073601} (\bibinfo {year}
  {2019})}\BibitemShut {NoStop}%
\bibitem [{\citenamefont {Ma}\ \emph {et~al.}(2011)\citenamefont {Ma},
  \citenamefont {Dai}, \citenamefont {Guo}, \citenamefont {Niu}, \citenamefont
  {Lu},\ and\ \citenamefont {Huang}}]{ma2011electronic}%
  \BibitemOpen
  \bibfield  {author} {\bibinfo {author} {\bibfnamefont {Y.}~\bibnamefont
  {Ma}}, \bibinfo {author} {\bibfnamefont {Y.}~\bibnamefont {Dai}}, \bibinfo
  {author} {\bibfnamefont {M.}~\bibnamefont {Guo}}, \bibinfo {author}
  {\bibfnamefont {C.}~\bibnamefont {Niu}}, \bibinfo {author} {\bibfnamefont
  {J.}~\bibnamefont {Lu}},\ and\ \bibinfo {author} {\bibfnamefont
  {B.}~\bibnamefont {Huang}},\ }\bibfield  {title} {\bibinfo {title}
  {Electronic and magnetic properties of perfect, vacancy-doped, and nonmetal
  adsorbed \ce{MoSe2}, \ce{MoTe2} and \ce{WS2} monolayers},\ }\href@noop {}
  {\bibfield  {journal} {\bibinfo  {journal} {Phys.~Chem.~Chem.~Phys.~}\
  }\textbf {\bibinfo {volume} {13}},\ \bibinfo {pages} {15546} (\bibinfo {year}
  {2011})}\BibitemShut {NoStop}%
\bibitem [{\citenamefont {Island}\ \emph {et~al.}(2016)\citenamefont {Island},
  \citenamefont {Kuc}, \citenamefont {Diependaal}, \citenamefont
  {Bratschitsch}, \citenamefont {Van Der~Zant}, \citenamefont {Heine},\ and\
  \citenamefont {Castellanos-Gomez}}]{isla+16ns}%
  \BibitemOpen
  \bibfield  {author} {\bibinfo {author} {\bibfnamefont {J.~O.}\ \bibnamefont
  {Island}}, \bibinfo {author} {\bibfnamefont {A.}~\bibnamefont {Kuc}},
  \bibinfo {author} {\bibfnamefont {E.~H.}\ \bibnamefont {Diependaal}},
  \bibinfo {author} {\bibfnamefont {R.}~\bibnamefont {Bratschitsch}}, \bibinfo
  {author} {\bibfnamefont {H.~S.}\ \bibnamefont {Van Der~Zant}}, \bibinfo
  {author} {\bibfnamefont {T.}~\bibnamefont {Heine}},\ and\ \bibinfo {author}
  {\bibfnamefont {A.}~\bibnamefont {Castellanos-Gomez}},\ }\bibfield  {title}
  {\bibinfo {title} {Precise and reversible band gap tuning in single-layer
  \ce{MoSe2} by uniaxial strain},\ }\href@noop {} {\bibfield  {journal}
  {\bibinfo  {journal} {Nanoscale}\ }\textbf {\bibinfo {volume} {8}},\ \bibinfo
  {pages} {2589} (\bibinfo {year} {2016})}\BibitemShut {NoStop}%
\bibitem [{\citenamefont {Johari}\ and\ \citenamefont
  {Shenoy}(2012)}]{joha-shen12nano}%
  \BibitemOpen
  \bibfield  {author} {\bibinfo {author} {\bibfnamefont {P.}~\bibnamefont
  {Johari}}\ and\ \bibinfo {author} {\bibfnamefont {V.~B.}\ \bibnamefont
  {Shenoy}},\ }\bibfield  {title} {\bibinfo {title} {Tuning the electronic
  properties of semiconducting transition metal dichalcogenides by applying
  mechanical strains},\ }\href@noop {} {\bibfield  {journal} {\bibinfo
  {journal} {ACS~Nano}\ }\textbf {\bibinfo {volume} {6}},\ \bibinfo {pages}
  {5449} (\bibinfo {year} {2012})}\BibitemShut {NoStop}%
\bibitem [{\citenamefont {Horzum}\ \emph {et~al.}(2013)\citenamefont {Horzum},
  \citenamefont {Sahin}, \citenamefont {Cahangirov}, \citenamefont {Cudazzo},
  \citenamefont {Rubio}, \citenamefont {Serin},\ and\ \citenamefont
  {Peeters}}]{horz+13prb}%
  \BibitemOpen
  \bibfield  {author} {\bibinfo {author} {\bibfnamefont {S.}~\bibnamefont
  {Horzum}}, \bibinfo {author} {\bibfnamefont {H.}~\bibnamefont {Sahin}},
  \bibinfo {author} {\bibfnamefont {S.}~\bibnamefont {Cahangirov}}, \bibinfo
  {author} {\bibfnamefont {P.}~\bibnamefont {Cudazzo}}, \bibinfo {author}
  {\bibfnamefont {A.}~\bibnamefont {Rubio}}, \bibinfo {author} {\bibfnamefont
  {T.}~\bibnamefont {Serin}},\ and\ \bibinfo {author} {\bibfnamefont
  {F.}~\bibnamefont {Peeters}},\ }\bibfield  {title} {\bibinfo {title} {Phonon
  softening and direct to indirect band gap crossover in strained single-layer
  \ce{MoSe2}},\ }\href@noop {} {\bibfield  {journal} {\bibinfo  {journal}
  {Phys.~Rev.~B}\ }\textbf {\bibinfo {volume} {87}},\ \bibinfo {pages} {125415}
  (\bibinfo {year} {2013})}\BibitemShut {NoStop}%
\bibitem [{\citenamefont {Scalise}\ \emph {et~al.}(2012)\citenamefont
  {Scalise}, \citenamefont {Houssa}, \citenamefont {Pourtois}, \citenamefont
  {Afanas’~ev},\ and\ \citenamefont {Stesmans}}]{scal+12nr}%
  \BibitemOpen
  \bibfield  {author} {\bibinfo {author} {\bibfnamefont {E.}~\bibnamefont
  {Scalise}}, \bibinfo {author} {\bibfnamefont {M.}~\bibnamefont {Houssa}},
  \bibinfo {author} {\bibfnamefont {G.}~\bibnamefont {Pourtois}}, \bibinfo
  {author} {\bibfnamefont {V.}~\bibnamefont {Afanas’~ev}},\ and\ \bibinfo
  {author} {\bibfnamefont {A.}~\bibnamefont {Stesmans}},\ }\bibfield  {title}
  {\bibinfo {title} {Strain-induced semiconductor to metal transition in the
  two-dimensional honeycomb structure of \ce{MoS2}},\ }\href@noop {} {\bibfield
   {journal} {\bibinfo  {journal} {Nano~Res.~}\ }\textbf {\bibinfo {volume}
  {5}},\ \bibinfo {pages} {43} (\bibinfo {year} {2012})}\BibitemShut {NoStop}%
\bibitem [{\citenamefont {Neupane}\ \emph {et~al.}(2021)\citenamefont
  {Neupane}, \citenamefont {Tang}, \citenamefont {Nepal}, \citenamefont
  {Adhikari},\ and\ \citenamefont {Ruzsinszky}}]{neupane2021opening}%
  \BibitemOpen
  \bibfield  {author} {\bibinfo {author} {\bibfnamefont {B.}~\bibnamefont
  {Neupane}}, \bibinfo {author} {\bibfnamefont {H.}~\bibnamefont {Tang}},
  \bibinfo {author} {\bibfnamefont {N.~K.}\ \bibnamefont {Nepal}}, \bibinfo
  {author} {\bibfnamefont {S.}~\bibnamefont {Adhikari}},\ and\ \bibinfo
  {author} {\bibfnamefont {A.}~\bibnamefont {Ruzsinszky}},\ }\bibfield  {title}
  {\bibinfo {title} {Opening band gaps of low-dimensional materials at the
  meta-{GGA} level of density functional approximations},\ }\href@noop {}
  {\bibfield  {journal} {\bibinfo  {journal} {Phys.~Rev.~Mater.}\ }\textbf
  {\bibinfo {volume} {5}},\ \bibinfo {pages} {063803} (\bibinfo {year}
  {2021})}\BibitemShut {NoStop}%
\bibitem [{\citenamefont {Cheng}\ \emph {et~al.}(2020)\citenamefont {Cheng},
  \citenamefont {Jiang}, \citenamefont {Li}, \citenamefont {Zhang},
  \citenamefont {Hu}, \citenamefont {Xie}, \citenamefont {Liu},\ and\
  \citenamefont {Qi}}]{cheng2020using}%
  \BibitemOpen
  \bibfield  {author} {\bibinfo {author} {\bibfnamefont {X.}~\bibnamefont
  {Cheng}}, \bibinfo {author} {\bibfnamefont {L.}~\bibnamefont {Jiang}},
  \bibinfo {author} {\bibfnamefont {Y.}~\bibnamefont {Li}}, \bibinfo {author}
  {\bibfnamefont {H.}~\bibnamefont {Zhang}}, \bibinfo {author} {\bibfnamefont
  {C.}~\bibnamefont {Hu}}, \bibinfo {author} {\bibfnamefont {S.}~\bibnamefont
  {Xie}}, \bibinfo {author} {\bibfnamefont {M.}~\bibnamefont {Liu}},\ and\
  \bibinfo {author} {\bibfnamefont {Z.}~\bibnamefont {Qi}},\ }\bibfield
  {title} {\bibinfo {title} {Using strain to alter the energy bands of the
  monolayer \ce{MoSe2}: A systematic study covering both tensile and
  compressive states},\ }\href@noop {} {\bibfield  {journal} {\bibinfo
  {journal} {Appl.~Surf.~Sci.~}\ }\textbf {\bibinfo {volume} {521}},\ \bibinfo
  {pages} {146398} (\bibinfo {year} {2020})}\BibitemShut {NoStop}%
\bibitem [{\citenamefont {Yu}\ \emph {et~al.}(2015)\citenamefont {Yu},
  \citenamefont {Xiong}, \citenamefont {Eshun}, \citenamefont {Yuan},\ and\
  \citenamefont {Li}}]{yu2015phase}%
  \BibitemOpen
  \bibfield  {author} {\bibinfo {author} {\bibfnamefont {S.}~\bibnamefont
  {Yu}}, \bibinfo {author} {\bibfnamefont {H.~D.}\ \bibnamefont {Xiong}},
  \bibinfo {author} {\bibfnamefont {K.}~\bibnamefont {Eshun}}, \bibinfo
  {author} {\bibfnamefont {H.}~\bibnamefont {Yuan}},\ and\ \bibinfo {author}
  {\bibfnamefont {Q.}~\bibnamefont {Li}},\ }\bibfield  {title} {\bibinfo
  {title} {Phase transition, effective mass and carrier mobility of \ce{MoS2}
  monolayer under tensile strain},\ }\href@noop {} {\bibfield  {journal}
  {\bibinfo  {journal} {Appl.~Surf.~Sci.~}\ }\textbf {\bibinfo {volume}
  {325}},\ \bibinfo {pages} {27} (\bibinfo {year} {2015})}\BibitemShut
  {NoStop}%
\bibitem [{\citenamefont {Madsen}\ and\ \citenamefont
  {Singh}(2006)}]{mads-sing06cpc}%
  \BibitemOpen
  \bibfield  {author} {\bibinfo {author} {\bibfnamefont {G.~K.}\ \bibnamefont
  {Madsen}}\ and\ \bibinfo {author} {\bibfnamefont {D.~J.}\ \bibnamefont
  {Singh}},\ }\bibfield  {title} {\bibinfo {title} {Boltztrap. a code for
  calculating band-structure dependent quantities},\ }\href
  {https://doi.org/https://doi.org/10.1016/j.cpc.2006.03.007} {\bibfield
  {journal} {\bibinfo  {journal} {Comput.~Phys.~Commun.~}\ }\textbf {\bibinfo
  {volume} {175}},\ \bibinfo {pages} {67} (\bibinfo {year} {2006})}\BibitemShut
  {NoStop}%
\bibitem [{\citenamefont {Vast}\ \emph {et~al.}(2016)\citenamefont {Vast},
  \citenamefont {Sjakste}, \citenamefont {Kane},\ and\ \citenamefont
  {Trinite}}]{transport-book}%
  \BibitemOpen
  \bibfield  {author} {\bibinfo {author} {\bibfnamefont {N.}~\bibnamefont
  {Vast}}, \bibinfo {author} {\bibfnamefont {J.}~\bibnamefont {Sjakste}},
  \bibinfo {author} {\bibfnamefont {G.}~\bibnamefont {Kane}},\ and\ \bibinfo
  {author} {\bibfnamefont {V.}~\bibnamefont {Trinite}},\ }\bibinfo {title}
  {Electronic transport: Electrons, phonons and their coupling within the
  density functional theory},\ in\ \href
  {https://doi.org/https://doi.org/10.1002/9781118761793.ch2} {\emph {\bibinfo
  {booktitle} {Simulation of Transport in Nanodevices}}}\ (\bibinfo
  {publisher} {John Wiley \& Sons, Ltd},\ \bibinfo {year} {2016})\
  Chap.~\bibinfo {chapter} {2}, pp.\ \bibinfo {pages} {31--96},\ \Eprint
  {https://arxiv.org/abs/https://onlinelibrary.wiley.com/doi/pdf/10.1002/9781118761793.ch2}
  {https://onlinelibrary.wiley.com/doi/pdf/10.1002/9781118761793.ch2}
  \BibitemShut {NoStop}%
\bibitem [{\citenamefont {Harada}\ \emph {et~al.}(2014)\citenamefont {Harada},
  \citenamefont {Sato},\ and\ \citenamefont
  {Yokoyama}}]{harada2014computational}%
  \BibitemOpen
  \bibfield  {author} {\bibinfo {author} {\bibfnamefont {N.}~\bibnamefont
  {Harada}}, \bibinfo {author} {\bibfnamefont {S.}~\bibnamefont {Sato}},\ and\
  \bibinfo {author} {\bibfnamefont {N.}~\bibnamefont {Yokoyama}},\ }\bibfield
  {title} {\bibinfo {title} {Computational study on electrical properties of
  transition metal dichalcogenide field-effect transistors with strained
  channel},\ }\href@noop {} {\bibfield  {journal} {\bibinfo  {journal}
  {J.~Appl.~Phys.~}\ }\textbf {\bibinfo {volume} {115}},\ \bibinfo {pages}
  {034505} (\bibinfo {year} {2014})}\BibitemShut {NoStop}%
\bibitem [{\citenamefont {Yun}\ \emph {et~al.}(2012)\citenamefont {Yun},
  \citenamefont {Han}, \citenamefont {Hong}, \citenamefont {Kim},\ and\
  \citenamefont {Lee}}]{yun2012thickness}%
  \BibitemOpen
  \bibfield  {author} {\bibinfo {author} {\bibfnamefont {W.~S.}\ \bibnamefont
  {Yun}}, \bibinfo {author} {\bibfnamefont {S.}~\bibnamefont {Han}}, \bibinfo
  {author} {\bibfnamefont {S.~C.}\ \bibnamefont {Hong}}, \bibinfo {author}
  {\bibfnamefont {I.~G.}\ \bibnamefont {Kim}},\ and\ \bibinfo {author}
  {\bibfnamefont {J.}~\bibnamefont {Lee}},\ }\bibfield  {title} {\bibinfo
  {title} {Thickness and strain effects on electronic structures of transition
  metal dichalcogenides: \ce{2H-MX2} semiconductors (\ce{M}= \ce{Mo}, \ce{W};
  \ce{X}= \ce{S}, \ce{Se}, \ce{Te})},\ }\href@noop {} {\bibfield  {journal}
  {\bibinfo  {journal} {Phys.~Rev.~B}\ }\textbf {\bibinfo {volume} {85}},\
  \bibinfo {pages} {033305} (\bibinfo {year} {2012})}\BibitemShut {NoStop}%
\bibitem [{\citenamefont {Chang}\ \emph
  {et~al.}(2014{\natexlab{b}})\citenamefont {Chang}, \citenamefont {Register},\
  and\ \citenamefont {Banerjee}}]{chang2014ballistic}%
  \BibitemOpen
  \bibfield  {author} {\bibinfo {author} {\bibfnamefont {J.}~\bibnamefont
  {Chang}}, \bibinfo {author} {\bibfnamefont {L.~F.}\ \bibnamefont
  {Register}},\ and\ \bibinfo {author} {\bibfnamefont {S.~K.}\ \bibnamefont
  {Banerjee}},\ }\bibfield  {title} {\bibinfo {title} {Ballistic performance
  comparison of monolayer transition metal dichalcogenide \ce{MX2} (\ce{M}=
  \ce{Mo}, \ce{W}; \ce{X}= \ce{S}, \ce{Se}, \ce{Te}) metal-oxide-semiconductor
  field effect transistors},\ }\href@noop {} {\bibfield  {journal} {\bibinfo
  {journal} {J.~Appl.~Phys.~}\ }\textbf {\bibinfo {volume} {115}},\ \bibinfo
  {pages} {084506} (\bibinfo {year} {2014}{\natexlab{b}})}\BibitemShut
  {NoStop}%
\bibitem [{\citenamefont {Choudhuri}\ \emph {et~al.}(2019)\citenamefont
  {Choudhuri}, \citenamefont {Bhauriyal},\ and\ \citenamefont
  {Pathak}}]{chou+19cm}%
  \BibitemOpen
  \bibfield  {author} {\bibinfo {author} {\bibfnamefont {I.}~\bibnamefont
  {Choudhuri}}, \bibinfo {author} {\bibfnamefont {P.}~\bibnamefont
  {Bhauriyal}},\ and\ \bibinfo {author} {\bibfnamefont {B.}~\bibnamefont
  {Pathak}},\ }\bibfield  {title} {\bibinfo {title} {Recent advances in
  graphene-like {2D} materials for spintronics applications},\ }\href@noop {}
  {\bibfield  {journal} {\bibinfo  {journal} {Chem.~Mater.~}\ }\textbf
  {\bibinfo {volume} {31}},\ \bibinfo {pages} {8260} (\bibinfo {year}
  {2019})}\BibitemShut {NoStop}%
\bibitem [{\citenamefont {Krumland}\ \emph {et~al.}(2024)\citenamefont
  {Krumland}, \citenamefont {Velja},\ and\ \citenamefont
  {Cocchi}}]{krumland2024quantum}%
  \BibitemOpen
  \bibfield  {author} {\bibinfo {author} {\bibfnamefont {J.}~\bibnamefont
  {Krumland}}, \bibinfo {author} {\bibfnamefont {S.}~\bibnamefont {Velja}},\
  and\ \bibinfo {author} {\bibfnamefont {C.}~\bibnamefont {Cocchi}},\
  }\bibfield  {title} {\bibinfo {title} {Quantum dots in transition metal
  dichalcogenides induced by atomic-scale deformations},\ }\bibfield  {journal}
  {\bibinfo  {journal} {ACS~Photon.}\ }\href
  {https://doi.org/10.1021/acsphotonics.3c01470} {10.1021/acsphotonics.3c01470}
  (\bibinfo {year} {2024})\BibitemShut {NoStop}%
\end{thebibliography}

%

\end{document}